\documentclass[aps,nofootinbib,notitlepage]{revtex4}
\usepackage{graphicx}
\usepackage{xcolor}
\usepackage{stmaryrd}
\usepackage{amstext}
\usepackage{etoolbox}
\usepackage{makeidx}
\usepackage{amsfonts}
\usepackage{amsmath,amssymb,epsfig}
\usepackage{amscd}
\usepackage{amsthm}
\usepackage{mathrsfs}
\usepackage{dsfont}
\usepackage[applemac]{inputenc}
\usepackage[english]{babel}
\usepackage{enumitem} 
\usepackage{subfigure}
\usepackage{hyperref}
\usepackage{blindtext}
\usepackage{verbatim}
\usepackage{cancel}
\usepackage{cases}
\usepackage{tensor}

\definecolor{crimson}{rgb}{0.7, 0.08, 0.24}

\newcommand{\be}{\begin{equation}}
	
	\newcommand{\ee}{\end{equation}}
\def\beqa{\begin{eqnarray}}
	\def\eeqa{\end{eqnarray}}
\def\bean{\begin{eqnarray*}}
	\def\eean{\end{eqnarray*}}
\def\nn{\nonumber}

\renewenvironment{thebibliography}[1]
{\section*{References}\frenchspacing\small
	\begin{list}{[\arabic{enumi}]}
		{\usecounter{enumi}\parsep=2pt\topsep 0pt
			\settowidth{\labelwidth}{[#1]}
			\leftmargin=\labelwidth\advance\leftmargin\labelsep
			\rightmargin=0pt\itemsep=1pt\sloppy}}{\end{list}}

\numberwithin{equation}{section}
\usepackage[normalem]{ulem}
\def\la{\langle}
\def\ra{\rangle}
\newcommand{\bra}[1]{\la {#1}|}
\newcommand{\ket}[1]{|{#1}\ra}

\newcommand{\R}{\mathbb{R}}
\newcommand{\C}{\mathbb{C}}

\usepackage{graphicx}
\usepackage{epstopdf}
\usepackage{amsthm}
\usepackage{tocvsec2}
\usepackage{enumerate}
\usepackage[T1]{fontenc}
\usepackage{color}
\usepackage{xcolor}
\usepackage{mathtools}
\usepackage{bbm}
\usepackage{braket}
\usepackage{tcolorbox}
\usepackage{natbib}
\usepackage{graphicx}

\usepackage{pgfplots}

\pgfplotsset{compat = newest}

\definecolor{darkgreen}{rgb}{0.0, 0.2, 0.13}
\definecolor{darkred}{rgb}{0.55, 0.0, 0.0}
\definecolor{carrotorange}{rgb}{0.93, 0.57, 0.13}
\usepackage{hyperref}
\hypersetup{
	colorlinks=true,
	linkcolor=blue,
	filecolor=magenta,
	urlcolor=cyan,
}
\usepackage[squaren,Gray]{SIunits}

\newcommand{\van}{\scriptstyle}

\newcommand{\bee}{\nopagebreak[3]\begin{equation*}}
	\newcommand{\eee}{\end{equation*}}

\newcommand{\ba}{\nopagebreak[3]\begin{eqnarray}}
	\newcommand{\ea}{\end{eqnarray}}
\DeclareFontFamily{U}{rsfs}{}         
\DeclareFontShape{U}{rsfs}{m}{n}{<5> rsfs5 <6><7> rsfs7          %
	<8><9><10><10.95><12><14.4><17.28><20.74><24.88> rsfs10}{}     %
\DeclareMathAlphabet{\mathfs}{U}{rsfs}{m}{n}                     %
\newcommand{\mfs}[1]{\mathfs {#1}}

\newcommand{\sH}{{\mfs H}}
\newcommand{\sL}{{\mfs L}}

\newcommand{\sI}{{\mfs I}}

\usepackage{braket}

\usepackage{etoolbox}

\begin{document}

\title{\textbf{\textsf{Discreteness Unravels the Black Hole Information Puzzle: Insights from a Quantum Gravity Toy Model}}\vspace{0.25cm}}

\author{Alejandro Perez}
\affiliation{{Aix Marseille Universit\'e, Universit\'e de Toulon, CNRS, CPT, Marseille, France}}
\date{\today}

\author{Sami Viollet}
\affiliation{{Aix Marseille Universit\'e, Universit\'e de Toulon, CNRS, CPT, Marseille, France}}

\begin{abstract}
	
\end{abstract}
\pacs{98.80.Es, 04.50.Kd, 03.65.Ta}
\begin{abstract}

The black hole information puzzle can be resolved if two conditions are met. Firstly, if the information of what falls inside a black hole remains encoded in degrees of freedom that persist after the black hole completely evaporates. These degrees of freedom should be capable of purifying the information. Secondly, if these purifying degrees of freedom do not significantly contribute to the system's energy, as the macroscopic mass of the initial black hole has been radiated away as Hawking radiation to infinity. The presence of microscopic degrees of freedom at the Planck scale provides a natural mechanism for achieving these two conditions without running into the problem of the large pair-creation probabilities of standard remnant scenarios. In the context of Hawking radiation, the first condition implies that correlations between the {\em in} and {\em out} Hawking partner particles need to be transferred to correlations between the {\em microscopic degrees of freedom} and the {\em out} partners in the radiation. This transfer occurs dynamically when the {\em in} partners reach the singularity inside the black hole, entering the UV regime of quantum gravity where the interaction with the microscopic degrees of freedom becomes strong.
The second condition suggests that the conventional notion of the vacuum's uniqueness in quantum field theory should fail when considering the full quantum gravity degrees of freedom. In this paper, we demonstrate both key aspects of this mechanism using a solvable toy model of a quantum black hole inspired by loop quantum gravity.
\end{abstract}
\maketitle

\section{Motivation}

The discussion surrounding Hawking's information puzzle  in black hole formation and evaporation \cite{Hawking:1976ra} holds significant importance as it serves as a testing ground for ideas aimed at establishing a coherent framework for quantum gravity. Its resolution becomes particularly crucial as it requires the description of dynamics within the strong quantum gravitational regime near the interior singularity. This applies to both holographic-like scenarios \footnote{Here we use this terminology to designate various approaches based on the so-called holographic principle loosely defined by Bekenstein's idea that number of degrees of freedom inside a region are bounded by area in some way or another. This has led to various standpoints which do not necessarily agree in their details but all share a certain  common ground. Examples range from works that give a fundamental status to Bousso type of entropy bounds \cite{Bousso:2002ju}, works in the constext of the ADS-CFT correspondence \cite{Maldacena:1997re}, to t'Hooft's \cite{tHooft:1996rdg} and Susskind's ideas on complementarity \cite{Susskind:1993if}.}, where black hole entropy is a measure of the number of internal states of the black hole, and non-holographic scenarios, where the internal degrees of freedom of a black hole are not limited by the area of the horizon.

In the first case, the objective is to reproduce the Page curve \cite{Page:1993wv} and elucidate the mechanism by which information escapes from the horizon while the black hole remains macroscopic. Even when the geometry near the horizon remains semiclassical in such regime, any convincing resolution must involve the strong quantum gravity regime as,  due to the no-cloning property of quantum information,  it should account for the fact that no information actually reaches the near the singularity region.  {Perhaps with the exception of the fuzzball scenario} \cite{Mathur:2005zp}, holographic approaches frequently dismiss the question of the quantum dynamics near the singularity as if it were a non-existent matter \cite{Almheiri:2020cfm, Almheiri:2019hni} \footnote{This was the clear position of representatives of different
  approaches to the problem during a panel discussion in a recent online
  meeting of the International Quantum Gravity Society in 2021 including N.
  Engelhardt, S. Giddings, A. Perez, and A., S. Raju., 2021.}. In the second case, a perspective in which this work is framed, it is crucial to dynamically depict the fate of the internal degrees of freedom as they evolve towards the strong Planckian regime near the singularity since they are the carriers of the information that preserve unitarity\cite{Perez:2022jlm, Perez:2017cmj, Amadei:2019wjp}.  

An intriguing possibility arises in the second scenario where, following the emission of {most} of its mass through Hawking radiation, a black hole could potentially leave behind a Planck mass remnant. This object would contain an immense number of internal states that are correlated with the Hawking particles, thereby preserving the purity of the initial state. {This proposal, known as the `remnant scenario', is subject to several objections}. One major concern is the existence of particle-like entities with high entropy, which would give rise to significant pair production amplitudes for these exotic objects which would impact quantum processes at low energies (see \cite{Giddings:1993km} and references therein). The perceived issue stems from the assumption that remnants can be effectively described in the framework of effective field theory, a premise that becomes highly questionable when considering the extremely vast internal universes that lie beyond the horizons of old black holes from a general relativity perspective \cite{Christodoulou:2016tuu}.

Another proposal, known as Wheeler's bag-of-gold idea \cite{wheeler1964relativity}, is closely related, although it disregards the objection against effective field theory for the reasons mentioned earlier. Here the remnant is presumed to harbor the purifying degrees of freedom within a vast internal 'universe.' These degrees of freedom would gradually escape into the nearly flat external spacetime as extremely low-energy particles after the black hole has completed its evaporation. However, there are significant challenges associated with this proposal. If a black hole begins with a macroscopic mass $M$ and undergoes evaporation through standard Hawking radiation, following semiclassical expectations until its mass nears the Planck mass $m_p$, energy conservation dictates that the individual components of the purifying radiation emitted by the remnant possess energies smaller than $m_p^3/M^2$ for a remnant of Planckian dimensions.  At the superficial level it is already difficult to conceive how such long-wavelength particles could be emitted from an object that appears Planckian from an external perspective. At a deeper level, and despite some recent interest \cite{Ashtekar:2020ifw} in the context of loop quantum gravity, simple entropic considerations make such a proposal highly improbable within a theory that incorporates fundamental discreteness. Specifically, as mentioned before (and illustrated in Figure \ref{purif}), the Hawking partners that were initially correlated with the Hawking radiation are forced to hit the strong Planckian regime near the singularity residing within and thus interact with a new {enormous} number microscopic degrees of freedom: the atomistic structure of spacetime in such a theory. From an entropic standpoint, this would necessitate an exceedingly delicate conspiracy for these excitations to reorganize into extraordinarily long-wavelength particles within quantum field theory as they emerge from the remnant (for further discussion, see \cite{Perez:2022jlm}).

The literature explores several additional possibilities worth considering. One such possibility is that the internal universe of the remnant remains causally disconnected from the external universe where black hole evaporation occurs, without any leakage of particles. This scenario would effectively break unitarity for the outside world. Another possibility is that when gravity is incorporated, quantum mechanics may require modifications wherein unitarity ceases to be a fundamental law (as discussed in Surdarsky's perspective in \cite{Perez:2022jlm} and related references). Furthermore, it is conceivable that certain aspects of the semiclassical language employed to describe the problem are inadequate, and the final state of evaporation may preclude a meaningful mean field spacetime representation (as highlighted in \cite{Page:1979tc, Page:1993up} and more recently in \cite{Calmet:2023gbw, Calmet:2022bpo, Calmet:2023met, Calmet:2021cip}). In such cases, the non-local nature of quantum mechanics and gravity could blur the distinction between the inside and outside regions \cite{Raju:2020smc}. Presently, there are no substantial objections to these possibilities. To ensure clarity in presenting our specific proposal, we have chosen to relegate the discussion of these aspects to a secondary role, with the hope that readers can obtain a more comprehensive understanding from the contributions of other authors in this special issue. Perhaps, in the future, a consensus may emerge from the diverse perspectives that currently appear to conflict with one another. 

\begin{figure}[h!]
	{\includegraphics[height=9cm]{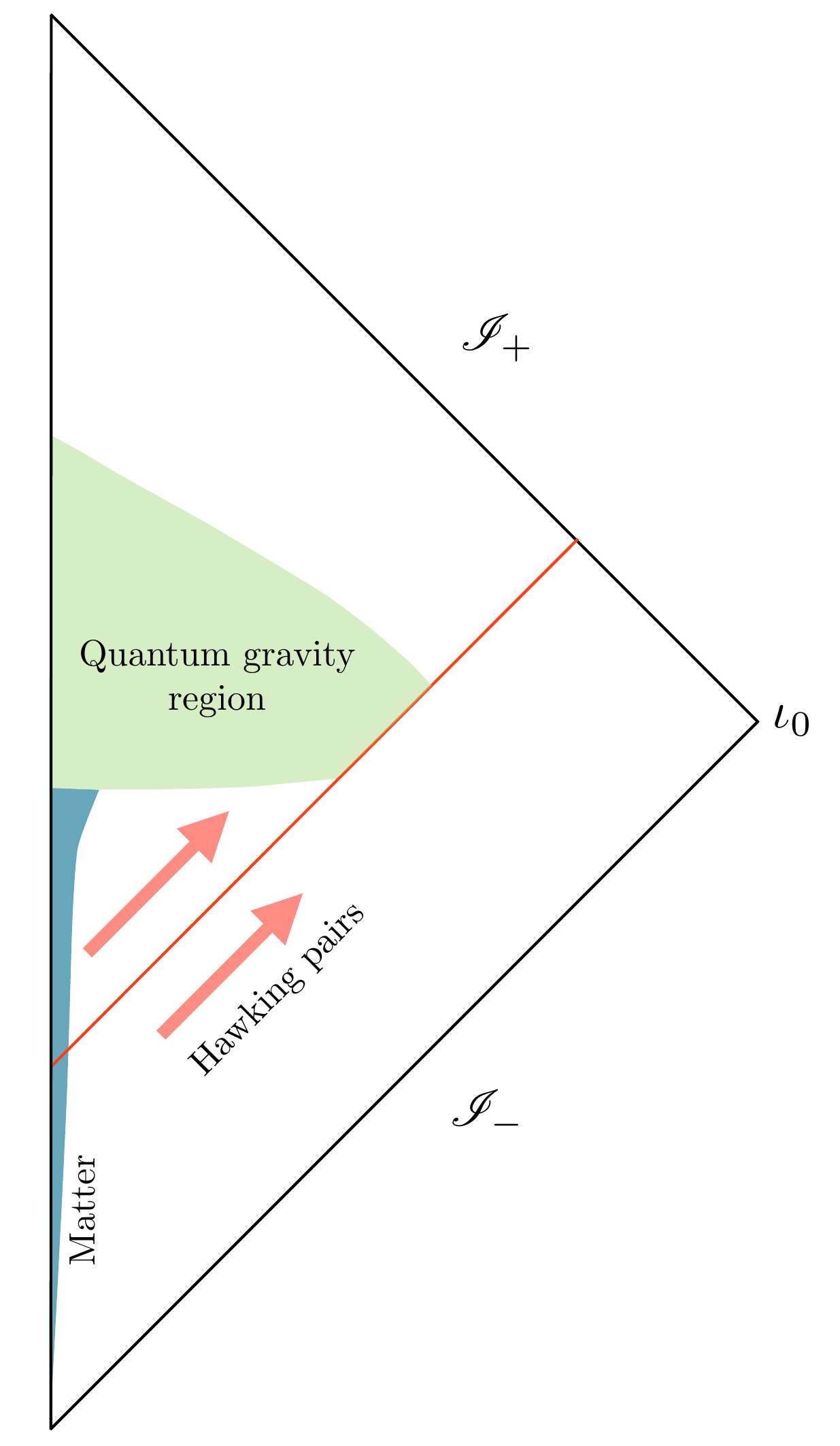}}
	\caption{The Ashtekar-Bojowald paradigm: the inside singularity is replaced by a quantum gravity region where no semiclassical or spacetime description is possible. Geometric observables (as well as matter degrees of freedom) cannot not be represented in terms of smooth field theoretic notions. Evolution across this region is expected to be well defined in the fundamental theory into a future where geometry and matter fields a well described (in a mean field sense) by a  nearly flat spacetime geometry in vacuum.}.
	\label{AA-MB}
\end{figure}

\begin{figure}[h!]
	{\includegraphics[height=9cm]{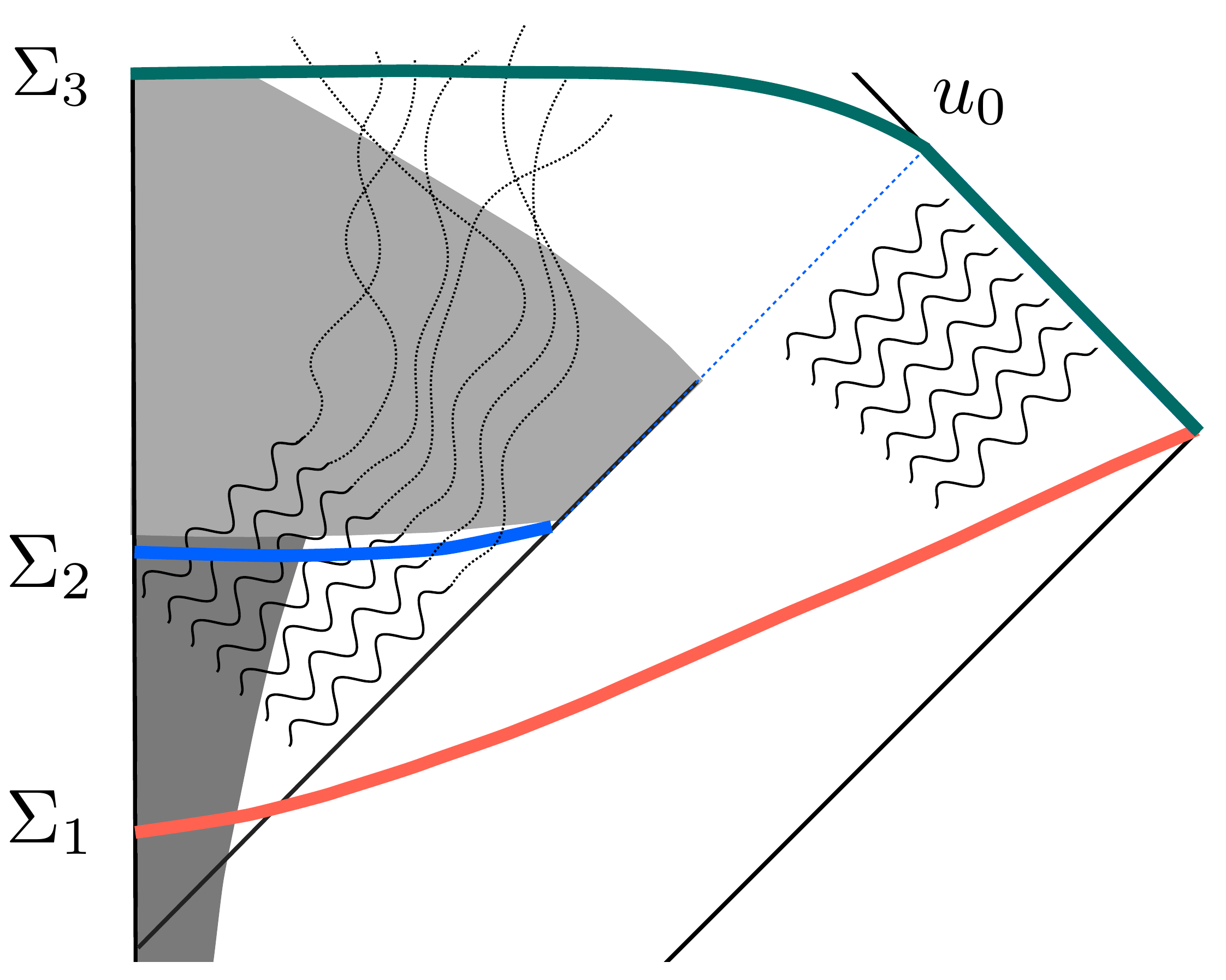}}
	\caption{Inside Hawking partners (initially maximally correlated with the Hawking radiation emitted to $\sI^+$) fall into the quantum gravity region where they are dynamically forced to interact with the microscopic granularity of quantum gravity. Near the singularity matter excitations are infinitely blue shifted into the Planckian regime. Such interactions transfer the entanglement from the initial field theoretic degrees of freedom in the Hawking partner to defects in the microscopic structure at the Planck scale. Defects emerge in the future of the quantum gravity region and their entanglement with the Hawking radiation purifies the final state. As the defects do not weight (the mean field flat geometry is degenerate) the process is compatible with energy conservation.}.
	\label{purif}
\end{figure}

Our perspective offers a straightforward and intuitive resolution of the paradox that aligns with other well-established descriptions of standard physical systems (like standard matter in statistical mechanics), in which information is degraded, yet not lost, due to the presence of microscopic granularity  \cite{Perez:2014xca}. The idea has been recently shown to be viable in certain models of analogue gravity \cite{Liberati:2019fse}. The proposal  is in sharp contrast with the standard remnant scenario, as the Planck-scale microscopic degrees of freedom purifying the Hawking radiation are not exclusively localized within a particle-like object but instead distributed throughout the fabric of spacetime after the complete evaporation of the black hole.  In this way our proposal avoids the objections raised against standard remnants. This possibility is based on the assumption that there is a quantum evolution across the singularity (in the context of loop quantum gravity the paradigm was first put forward in \cite{Ashtekar:2005cj} and includes more recent black-hole to white-hole transition models \cite{Han:2023wxg}).   The physical situation representing black hole formation and evaporation, of what we call the Ashtekar-Bojowald paradigm,  is illustrated by the Penrose diagram shown in Figure \ref{AA-MB} where the singularity is replaced by a region where no spacetime description is possible and a full quantum gravity treatment becomes mandatory. 

A pictorial representation of the scenario for purification that we propose is given in Figure \ref{purif}. Entanglement between the Hawking radiation going to infinity at $\sI^+$ and inside partners maintain unitarity in the semiclassical quantum field theory description that is expected to be valid while the black hole is macroscopic: in our illustration we assume this is so around the instant defined by the Cauchy surface $\Sigma_1$. Inside partners (wavy lines on the left of the horizon) fall into the quantum gravity region (what classically we call the singularity) and transfer their entanglement to defect-like Planckian degrees of freedom that cannot be described in an effective field theory framework (they are discrete excitations in an fundamentally discrete theory with no field theoretic analog). These are represented by dotted lines in the quantum gravity region. The evolution between $\Sigma_1$ and $\Sigma_3$ in the full quantum gravity theory includes the discrete fundamental degrees of freedom and is thus unitary. In contrast with fuzzballs, the amount of information that can be coded in these defects grows with the number of defects involved---which can be arbitrary depending on the internal extension of the black hole, e.g., one per Planck volume---and hence is by no means bounded by the area of the corresponding horizon. Information would seem to be lost in any effective description that neglects such microscopic defects. Now, for the scenario to be consistent with energy conservation, the Bondi mass of the system for retarded times after $u_0$ (retarded time of complete semiclassical evaporation) must be close to zero or about the Planck mass. This is possible if the macroscopic emergent geometry is degenerate, i.e., it corresponds to many microscopically inequivalent configurations. This is a feature expected to be present in certain approaches to quantum gravity. In the case of loop quantum gravity such property is precisely the one that explains the origin of black hole entropy in the semiclassical regime \cite{Krasnov:1996tb, Rovelli:1996dv, Ashtekar:1997yu, Ashtekar:2000eq} (for reviews see \cite{Perez:2017cmj, BarberoG:2015xcq} for more discussion of this specific feature see \cite{Perez:2022jlm})\footnote{ The existence of such fundamental granularity is also a potential source of interesting phenomenology: as argued in \cite{Amadei:2021aqd} diffusion effects could provide a natural paradigm for the generation of dark energy \cite{Josset:2016vrq, Perez:2017krv}, and (under suitable circumstances and with natural assumptions) provide the seeds for the inhomogeneities observed at the CMB in cosmology \cite{Amadei:2021aqd} or explain the origin of dark matter \cite{Barrau:2019cuo, Amadei:2021aqd}. }.

A fundamental description of the previous mechanism remains out of reach at the present stage of development of approaches to quantum gravity with Planckian granularity. However, it is possible to make the previous scenario concrete in simplified models.
The interior region $r< 2M$ of a Schwarzschild black hole of mass $M$ can be seen as a homogeneous anisotropic cosmological model where the $r$=constant surfaces (in the usual Schwarzschild coordinates) are Cauchy surfaces of homogeneity: any two arbitrary points can be connected along orbits of the isometry group who involves spacelike translations along the staticity Killing field $\xi=\partial_t$ and the rotations associated to spherical symmetry. Models with these isometries will be referred to as  Kantowski-Sachs (KS) models \cite{kantowski1966some}. They include not only the Schwarzschild black hole interior geometry (vacuum case) but also the Reissner-Nordstrom black hole interior geometry (in the Einstein-Maxwell case) and other solutions depending of the type of matter that one decides to couple to the system.
In a recent paper we showed that such Kantowski-Sachs models (with a massless scalar field coupling) define a natural toy model capturing aspects of the dynamics and back-reaction of matter near the singularity of realistic black holes that Hawking radiate and evaporate \cite{Perez:2023jrq}.  The model is quantized {\em \'a la loop} in that paper and it is shown that certain microscopic degrees of freedom (not relevant for coarse observers) arise as a consequence of peculiar representation of the algebra of observables used. Such, at first appearance, exotic representation is indeed the natural one when it comes to the construction of the background independent quantization of gravity in loop quantum gravity \cite{Lewandowski:2005jk}. The presence of microscopic degrees of freedom 
that remain hidden at low energies is an expected generic feature in the full quantum gravity regime \footnote{Even when the continuum limit remains an open issue in the framework of loop quantum gravity, the are various potential sources of 
micrsocopic degeneracy in the structure of the theory. For instance, consistent regularizations of the Hamiltonian constraint 
all seem to share the property of being insensitive to certain microscopic details of the quantum states that could code informations. These defect-like structures are related to the fundamental degeneracy of the quantum geometry at the microscopic level. There are studies showing that in the emergent low energy regime some degrees of freedom would decouple as they would appear as highly massive \cite{Dittrich:2021kzs, Dittrich:2022yoo, Borissova:2022clg}.}.

{The existence of such a hidden sector (which represents a sort
	of quantum hair) is central in the scenario for purification of the Hawking
	radiation in black hole evaporation proposed in \cite{Perez:2014xca}. In the present paper we use the simple black hole model recalled above to} illustrate the key ingredients of the general scenario: firstly we review how microscopic degrees of freedom arise in the quantization, secondly we show how quantum correlations between matter excitations and such microscopic degrees of freedom develop unavoidably during dynamical evolution. This suggests  that the microscopic physics does play an important role in discussions about the fate of unitarity in black hole formation and evaporation, and that any effective field theory approach to the unitarity  question that avoids confronting the details of the dynamics near the singularity misses a central aspect and thus cannot provide a complete picture of the resolution of the puzzle.    

The article is organized as follows. In Section \ref{KSM} we review the definition of the simplified model for the dynamics of matter and geometry inside of a spherically symmetric black hole. In this section we basically reproduce the results and aspects of the presentation of  \cite{Perez:2023jrq}. We give the definition of the mini-superspace model of interest and review its polymer quantization. In Section \ref{model} we use the simplified setup to construct a minimalistic model of the dynamics of a Hawking pair on our simplifies black hole setup. The back reaction of the in falling partner is suitably taken into account as it approaches the strong field regime near the singularity where the coupling between geometry and matter becomes most important. The outside partner dynamics is assumed to be trivial (idealizing the weak back reaction for wave packets evolving toward infinity; completely neglected in the framework of the Hawking effect) however the model suitably keeps track of the entanglement with the inside: the full dynamics is unitary. In Section \ref{resultats} we present the key results supporting our claims. We end with a discussion of the paper in Section \ref{disc}.

\section{The model}\label{KSM}

{We start by reviewing the model} presented in \cite{Perez:2023jrq} that we will use later on in this work to calculate the back reaction of a Hawking pair as one of the partners falls into the black hole singularity.
In a description on an $r$ equal constant slicing of the interior of a Schwarzschild black hole, a free test scalar excitation (with no angular momentum) is well approximated as it approaches the singularity by a translational invariant wave function, as  the expansion in the spacelike Killing direction $\xi=\partial_t$ diverges for $r\to 0$. This means that zero angular momentum test particles can be approximated by the type of excitations that can be accommodated in the dynamical framework of the KS cosmologies (at least in the sense of a near singularity approximation) \footnote{One can be quantitative about this intuition as follows:
 free test particles with four wave vector $k^a$ on the Schwarzschild background are associated with the conserved Killing energy ${\cal E}\equiv -k^a\xi_a$. We are assuming that the particle has zero angular momentum which implies that its wave function is already translational invariant in the directions transversal to $\xi^a$ on the $r$-slices. The wave function can only vary in the direction of the Killing $\xi^a$ and  the component of the physical momentum in this direction is given by 
 \be\label{lapu}
 p_\xi \equiv \frac{k^a\xi_a}{\sqrt{\xi\cdot\xi}}=-{\cal E}\sqrt{\frac{r}{2M-r}},
 \ee
 which vanishes in the limit $r\to 0$. The wave length of such a particle diverges, and thus particles without angular momentum are better and better represented by translational invariant excitations as one approaches the singularity.}.
This simple implication, deduced from the idealized notion of test particle, can be made more precise by looking at the analogous features of scalar field excitations (solutions of the Klein-Gordon equation). Indeed, the simplistic argument given here can be made precise as shown in \cite{Perez:2023jrq}.
Therefore, in what follows we will review the construction of a quantum model of the black hole interior that can accommodate such scalar excitations and their back reaction. Such model will set the stage where an idealized Hawking pair will be defined and the quantum information dynamical details, relevant for our discussion,  of will be computed. 

\subsection{Symmetry reduced covariant phase space} \label{ps}

Here we briefly report the construction of the model introduced in  \cite{Perez:2023jrq} (refer to that paper for more details). The first step towards canonical quantization is to define the Hamiltonian description of the model. It is well known that for a spherically symmetric and static spacetime, the line element can be written without any loss of generality as 
\begin{equation}\label{ansatz}
ds^2=-f(r)dt^2+h(r)dr^2+r^2d\Omega^2 \ .
\end{equation}
It follows that the Eintein-Hilbert action (with the appropriate boundary term that renders it differentiable: the usual Gibbons-Hawking-York boundary term) becomes
\be\label{ss}
S_{\rm geo}=\frac{1}{16\pi \ell_p^2}\left[\int_R d^4x\sqrt{-g}R + 2 \int_{\partial R} K\right] =\frac{\ell_0}{2\ell_p^2} \int dr \left(\sqrt{fh}+\sqrt{\frac{f}{h}} +\frac{\dot f r}{\sqrt{fh}}\right),\ee
where $\ell_p^2=G$ (in $\hbar=1$ units) and $\ell_0$ is a infra-red cut off in the Killing time $t$ necessary to make the action of the symmetry reduced model finite (for more details see \cite{Perez:2023jrq}).
From which we read the Lagrangian $\sL_{\textrm{\rm geo}}$ of the spacetime subsystem
\begin{equation}
\sL_{\rm geo}=\frac{\ell_0}{2\ell_p^2} \left(\sqrt{fh}+\sqrt{\frac{f}{h}}+\frac{r\dot f}{\sqrt{fh}}\right) .
\end{equation}
On the other hand, we will couple the system to a massless scalar field by adding the matter action
\be
S_m=-\frac{1}{2}\int_R d^4x\sqrt{-g}\partial_a\phi \partial^a\phi =-2\pi \ell_0 \int dr r^2 \dot \phi ^2\sqrt{\frac{f}{h}}.
\ee
The conjugate momenta to $f,h$ and $\phi$ are given by
\begin{equation}\label{conj m}
{\rm p}_f=\frac{\ell_0}{2\ell_p^2}\frac{r}{\sqrt{fh}} \quad , \quad {\rm p}_h=0 \quad \textrm{and} \quad {\rm p}_\phi=-4\pi r^2 \ell_0 \sqrt{\frac{f}{h}} \dot \phi,
\end{equation}
and the primary Hamiltonian, defined by $H=\dot f {\rm p}_f+\dot h {\rm p}_h+\dot \phi {\rm p}_\phi-\sL_{\phi}-\sL_{\rm geo}$, becomes
\begin{equation} \label{p H}
{\rm H}_1=-\frac{\ell_0}{2\ell_p^2} \frac{f(h+1)}{\sqrt{fh}}
-\frac{h{\rm p}_\phi^2}{8\pi r^2 \ell_0 \sqrt{fh}} \ .
\end{equation}
From the expression of the conjugate momenta \eqref{conj m}  we identify the constraints \begin{equation}\label{constraints}
\xi\equiv {\rm p}_f-\frac{\ell_0}{2\ell_p^2} \frac{r}{\sqrt{fh}}=0 \quad \textrm{and} \quad {\rm p}_h=0,
\end{equation}
and the secondary Hamiltonian 
\begin{equation}\label{s H}
	{\rm H}_2={\rm H}_1+\lambda \xi + \eta {\rm p}_h \ ,
\end{equation}
where $\lambda$ and $\eta$ are Lagrange multipliers. One can show that the stability of the two constraints \eqref{constraints} can be ensured by fixing  the associated Lagrange multipliers, i.e.,  the constraints \eqref{constraints} are second class and can be explicitly solved leading to  
\begin{equation}\label{central}
	{\rm p}_h=0 \quad \textrm{and} \quad h= \frac{\ell^2_0}{4\ell_p^4}\frac{r^2}{f{\rm p}_f^2} .
\end{equation}
Thus, the secondary Hamiltonian \eqref{s H} reduces to
\begin{equation}\label{wewe}
	{\rm H}_2=-\frac{1}{r}\left(f {\rm p}_f+\frac{1}{16\pi \ell_p^2} \frac{{\rm p}_{\phi}^2}{f {\rm p}_f}+\frac{\ell^2_0}{4\ell_p^4} \frac{r^2}{{\rm p}_f} \right)\ .
\end{equation}
The previous Hamiltonian encodes the KS dynamics of geometry coupled to a massless scalar field.
The relevant solutions for physical applications correspond to small departures from the vacuum Schwarzschild solutions representing macroscopic black holes with scalar field perturbation falling inside. The system is simplified further by focusing on, what we call, the 
{deep interior region} $r\ll M$ where $M$ is the mass scale defined by the corresponding black hole solution perturbed by the presence of matter. It is in this regime where the solutions of the KS system faithfully describe the dynamics of a spherically symmetric scalar perturbation (representing for instance a Hawking particle) as it falls towards the interior singularity.  It is in this regime that the KS Hamiltonian evolution given by \eqref{wewe} matches,  the test-field evolution (the Klein-Gordon solutions on the Schwarzschild background fixed non dynamical background) and incorporates, as a simplified model, aspects of the back-reaction that are expected to become more important as one approaches the singularity.  As shown in \cite{Perez:2023jrq}, the simplification simply amounts to dropping the last term in the previous Hamiltonian, namely
\begin{equation}\label{HDI}
	{\rm H}_{\rm di}=-\frac{1}{r}\left(f {\rm p}_f+\frac{1}{16\pi \ell_p^2} \frac{{\rm p}_{\phi}^2}{f {\rm p}_f}\right).
\end{equation}
This toy theory reflects the dynamics of the leading order in an expansion near $r=0$. 
 In order to recover the structure of a gauge theory, with a clear analogy with the full theory of LQG, it is convenient to `reparametrize' the system by promoting the area radius $r$ to a degree of freedom with conjugate momentum ${\rm p}_r$ and add a scalar constraint ${\rm C}={\rm p}_r-{\rm H}_2=0$. Introducing the {\em deep interior variables} 
\begin{equation}\label{nsv}
	m=-f {\rm p}_f \ \ \ \textrm{and} \ \ \ {\rm p}_m=-\log(-f) \ ,
\end{equation}
and adopting the area $a$ of the surfaces of constant $r$ as time
\begin{equation}
	a=4\pi r^2 \quad \textrm{and} \quad {\rm p}_a=\frac{{\rm p}_r}{8\pi r} \ ,
\end{equation}
one has
\begin{align*} \label{classical ccr}
	\{m,{\rm p}_m\}&=1 \ , \quad \{a,{\rm p}_a\}=1
\ ,
\end{align*}
and
\begin{equation}\label{classical ccr mat}
	\{\phi,{\rm p}_\phi\}=1,
\end{equation}
with all the other Poisson brackets equal to zero.
The Hamiltonian constraint describing the deep interior regime is 
\begin{equation} \label{Hc 2}
\boxed{{\rm C}_a={\rm p}_a+\frac{1}{2a}\left(m+\frac{{\rm p}_{\phi}^2}{16\pi \ell_p^2 m}\right) \approx 0 .}
\end{equation}
The previous constraint defines the classical dynamical equation of the model. In the quantum theory---and in the loop representation that mimics the one used in the full LQG framework---the area variable will evolve in discrete steps. A side effect of such discreteness is the appearance of discrete quantum hair:  black hole states are labelled by macroscopic quantum numbers corresponding to the eigenvalues of ${\rm p}_\phi$ and $m$, but also by a microscopic quantum number $\epsilon$ (quantum hair).

\subsection{Sketch of the Schroedinger quantization} \label{q geo}

For comparison, let us start by reviewing the Schroedinger quantization of the system presented in the last section.
In the standard Schroedinger representation one would quantize the phase space of Section \ref{ps} by promoting the variables $a, m, {\rm p}_a, {\rm p}_m$ to self adjoint operators
\ba\!\!\!\!\!\!\!\!	\begin{array}{ccc}
	&& \widehat m\ \psi(m, {\rm p}_\phi, a) = m \psi(m, {\rm p}_\phi, a), \\ && \widehat {\rm p}_{m} \psi(m, {\rm p}_\phi, a)=-i{\partial_m \psi(m, {\rm p}_\phi, a)}, \end{array}\ 
	\begin{array}{ccc} && \widehat a\ \psi(m, {\rm p}_\phi, a)=a\psi(m, {\rm p}_\phi, a),  \\ 
	  && \widehat{{\rm p}}_a \psi(m, {\rm p}_\phi, a)=-i {\partial_a \psi(m, {\rm p}_\phi, a)},\end{array} \  
	\begin{array}{ccc} && \widehat {\rm p}_\phi \psi(m, {\rm p}_\phi, a)={\rm p}_\phi \psi(m, {\rm p}_\phi, a),  \\ 
	  && \widehat{\phi}\ \psi(m, {\rm p}_\phi, a)=i {\partial_{{\rm p}_\phi} \psi(m, {\rm p}_\phi, a)},\end{array} \nn
\ea
in the kinematical Hilbert space is $\sH_{\rm S}={\sL}^2(\mathbb{R}^3) $, equipped with the usual inner product
\begin{equation} \label{usual ip}
	\langle \psi_1 , \psi_2 \rangle = \int_{-\infty}^{+\infty}\int_{-\infty}^{+\infty} \int_{-\infty}^{+\infty}   \overline{\psi_1(m, {\rm p}_\phi, a)} \psi_2(m, {\rm p}_\phi, a)  dm d{\rm p}_\phi  da  ,
\end{equation}
where we have chosen the momentum representation for the scalar field for convenience (as ${\rm p}_{\phi}$ is one of the constants of motion of the system).
Eigenstates of the $\widehat a$ operator are interpreted as distributions (they are not in the Hilbert space) and one usually writes
\be
\widehat a \ket{a}=a \ket{a}
\ee
with $a\in \R$ and form an orthonormal basis 
\be\label{kikin}
\braket{a,a'}=\delta({a,a'}).
\ee
The dynamics is imposed by solving the Hamiltonian constraint (\ref{Hc 2}) which, in the present representation, takes the precise form of a Schroedinger equation in the area variable $a$, namely 
\be\label{SE}
\left[-i\hbar \frac{\partial}{\partial a}+\frac{1}{2a}\left(m+\frac{{\rm p}_{\phi}^2}{16\pi \ell_p^2 m}\right)\right]\psi(m, {\rm p}_\phi, a)=0.
\ee
As usual, solutions of the constraint are certainly not square integrable in the $a$-direction, thus physical states are outside of the kinematical Hilbert. The physical Hilbert space is defined as the space of square integrable functions of $m$ and ${\rm p}_{\phi}$ at fixed time $a$---$\sH_{\rm phys}={\sL}^2(\mathbb{R}^2)$---with inner product
\be
\label{usual phys ip}
	\langle \psi_1(a), \psi_2(a) \rangle_{\rm phys} = \int_{-\infty}^{+\infty} \int_{-\infty}^{+\infty}   \overline{\psi_1(m, {\rm p}_\phi, a)} \psi_2(m, {\rm p}_\phi, a)  dm d{\rm p}_\phi ,
\ee 
which is preserved, i.e. it is independent of $a$,  by the Schroedinger equation (evolution is unitary in $a$).

Two important remarks are in order: 
First note that we are formulating in detail the dynamics of the system in the near singularity approximation. The physical reason for this is that (as argued previously) it is only in this approximation that the system can be compared with a (spherically symmetric) black hole with spherically symmetric excitations falling inside. A side gain is also the simplification of the dynamics which will allow us for a simpler quantization and the analysis of the possibility of a well defined dynamics across the singularity when we undergo the LQG inspired quantization. One could however consider the quantization of the minisuperspace system without the near singularity approximation. In that case one would need to write a Schroedinger equation using the Hamiltonian \eqref{wewe}, now genuinely time-dependent ($r$-dependent), for which unitary evolution would involve path ordered exponentials (as the Hamiltonian does not commute with itself at different $r$ values). In addition one would need to work with either $r,f, {\rm p}_r, {\rm p}_f$ variables or $a,f, {\rm p}_a, {\rm p}_f$ variables without the luxury of the simplifications introduced by the use of the near singularity variables \eqref{nsv}.    

 \subsection{The loopy quantization}\label{looloo}
 
We define now a representation of the phase space variables that incorporates a key feature of the full theory 
of LQG: the area quantization. This representation closely mimics the structure of the quantum theory in the fundamental theory in such a way that the area variable $a$ acquires a discrete spectrum. Mathematically, this is achieved by replacing the $\sL^2$ structure of the inner product in the variable ${\rm p}_a$ by the inner product of the Bohr compactification of the ${\rm p}_a$ phase space dimension. 
This is analogous to what is done in the full theory of loop quantum gravity that uses the Ashtekar-Barbero connection-variables \cite{Ashtekar:1986yd,Barbero:1994ap} as the starting point. There, the connection is not represented as a fundamental operator but only its holonomy (an exponentiated version of the connection in essence) in the definition of the Hilbert space \cite{Lewandowski:2005jk}.  More precisely one replaces the kinematical inner product in the Schroedinger representation \eqref{usual ip} by
\begin{equation} \label{bohr ip}
\langle \psi_1 , \psi_2 \rangle  = \lim_{\Delta \rightarrow +\infty} \frac{1}{2\Delta}\int_{-\Delta}^{+\Delta}  \left(\int_{-\infty}^{+\infty} \overline{\psi_1(m, {\rm p}_\phi, {\rm p}_a)} \psi_2(m, {\rm p}_\phi, {\rm p}_a)  dm d{\rm p}_\phi  \right)d{\rm p}_a \ .
\end{equation}
With this inner product, periodic functions of ${\rm p}_a$ with arbitrary period are normalizable and the conjugate $a$-representation acquires the property of discreteness in a way that closely mimics the structure of the fundamental  theory of loop quantum gravity \cite{Ashtekar:2011ni}. In particular eigenstates of $
\widehat a$ exist 
\be
\widehat a \ket{a}=a \ket{a}
\ee
with $a\in \R$. These states form an orthonormal basis with inner product 
\be\label{pipolo}
\braket{a,a'}=\delta_{a,a'},
\ee
in contrast with \eqref{kikin}.
Discreteness of the spectrum of $\widehat a$ comes at the prize of changing the kinematical Hilbert space structure in a way that {prevents the infinitesimal translation operator $\widehat {\rm p}_a$ from existing} (this is because the inner product \eqref{bohr ip} is not weakly continuous under the action of translations: according to \eqref{pipolo} an infinitesimally translated state is orthogonal to the original state). Instead only finite translations (quasi periodic functions of ${\rm p}_a$) can be represented as unitary operators in the polymer Hilbert space. Their action on the $a$-basis is given by
\begin{equation}\label{defidefi}
	\widehat{e^{i\lambda {\rm p}_a}}\psi(m, {\rm p}_\phi, a)=\psi(m, {\rm p}_\phi,a+\lambda \ell_p).
\end{equation}
Eigenstates of the finite translations (or shift operators)  exist and are given by wave functions supported on discrete $a$-lattices. Namely,
\be \label{cocuna} \psi_{k,\epsilon}(a)\equiv \left\{ \begin{array}{ccc} &\exp (i k a)& \ \ \ {\rm if} \ \ \ a\in \Gamma_{\epsilon,\lambda}\equiv \{(\epsilon+n\lambda)\ell_p^2\in \R \}_{n\in \mathbb{Z}} \\ &0& \ \ \ {\rm otherwise}
\end{array}\right.\ee
where  the parameter $\epsilon\in [0,\lambda)\in \R$. The discrete lattices denoted $\Gamma_{\epsilon,\lambda}$ are the analog of the spin-network graphs in LQG with the values of $a$ on lattice sites the analog of the corresponding spin labels.  With all this one has  (using \eqref{defidefi}) that
\be
\widehat{e^{i\lambda {\rm p}_a}}\psi_{k,\epsilon}(a)= {e^{i\lambda k}} \psi_{k,\epsilon}(a).
\ee
Note that, unlike the Schroedinger representation where the eigen-space of the momentum operator is one dimensional, the eigen-spaces of the translation operator (labelled by the eigenvalue ${e^{i\lambda k}}$) are infinite dimensional and non separable. This is explicit from the independence of the eigen-values of the continuous parameter $\epsilon\in [0, \lambda)$ labelling eigenstates. Such huge added degeneracy in the spectrum of the shift operators is a general feature of the polymer representation. We will show that this degeneracy can show up in Dirac observables of central physical importance such as the mass operator in Section \ref{degeM}. 

\subsection{Quantum dynamics}
In the full theory a new perspective on the regularization issue has been introduced 
motivated by the novel mathematical notion of generalized gauge covariant Lie derivatives \cite{Ashtekar:2020xll} and their geometric interpretation allowing for the introduction of a natural regularization (and subsequent) anomaly free quantization of the Hamiltonian constraint \cite{Varadarajan:2022dgg}. Even when the procedure  does not eliminate all ambiguities of quantization (choices are available in the part of the quantum constraint responsible for propagation \cite{ale-madhavan}), the new technique reduces drastically some of them in the part of the Hamiltonian that is more stringently constrained by the quantum algebra of surface deformations.  

{{The primary focus of our discussion is to highlight the similarity in the impact of the analogous process when applied to our symmetry reduced Hamiltonian. As our classical Hamiltonian constraint features linearity in the variable ${\rm p}_a$, whose corresponding Hamiltonian vector field has a clear geometric interpretation of infinitesimal translations in $a$ which in the quantum theory can only be represented by finite translations or shifts. }}  One makes the replacement
\be \lambda \widehat {\rm p}_a \ \ \ \longrightarrow \ \ \ \widehat{e^{i\lambda {\rm p}_a}}.\ee
{{Now in order to maintain geometric compatibility with the Schroedinger equation one exponentiates the second term in the classical Hamiltonian \eqref{Hc 2} obtaining the well-known unitary evolution operator producing finite area evolution. }} The quantum constraint preserving the geometric consistency with the Schroedinger equation is
\be\label{dydy}
{\underbrace{\exp({i\lambda {\rm p}_a})}_{\begin{array}{ccc} \rm \van finite \ areatime \\ \rm \van translation\end{array}}\ket{\psi} -  \overbrace{\exp\left({\frac{i}{2} \log\left(\frac{a+\lambda \ell^2_p}{ a}\right) \left(m+\frac{{\rm p}_{\phi}^2}{16\pi \ell_p^2 m}\right)}\right) }^{\begin{array}{ccc} \rm \van finite \ areatime \\ \rm \van unitary \ evolution \ operator\end{array}}}\ \ket{\psi} =0, 
\ee 
whose action is well defined in the polymer representation and whose solutions are easily found (by acting on the left with $\bra{m,{\rm p}_\phi,a}$) to be wave functions satisfying the discrete dynamics given by
\be\label{step}
\psi(m,{\rm p}_\phi, a+\lambda \ell_p^2)
=e^{\frac{i}{2} \log\left(\frac{a+\lambda \ell^2_p}{ a}\right) \left(m+\frac{{\rm p}_{\phi}^2}{16\pi \ell_p^2 m}\right)}  
\psi(m,{\rm p}_\phi, a).
\ee
The physical Hilbert space is defined via the usual inner product at fixed (discrete) time $a$ via
\be
\label{phys ip}
	\langle \psi_1(a), \psi_2(a) \rangle_{\rm phys} = \int_{-\infty}^{+\infty} \int_{-\infty}^{+\infty}   \overline{\psi_1(m, {\rm p}_\phi, a)} \psi_2(m, {\rm p}_\phi, a)  dm d{\rm p}_\phi ,
\ee   
which is independent of the lattice sites  as required (a property identified with the unitarity of the dynamics generated by the quantum constraint). More precisely the physical inner product is a constant of the quantum motion associated to the full history represented by the lattice $\Gamma_{\epsilon, \lambda}$ as implied by
unitarity.  Explicitly one has
\be
\langle \psi_1(a), \psi_2(a) \rangle_{\rm phys}= \langle \psi_1(a+\lambda), \psi_2(a+ \lambda) \rangle_{\rm phys}\ \ \ \forall \ \ \ a\in \Gamma_{\epsilon, \lambda}.
\ee
Ambiguities of regularization that are usually associated with the polymerization procedure are thus completely absent in this model. The reason is the linear dependence of the Hamiltonian constraint in the polymerized variable which allows for a regularization fixed by the geometric interpretation of the classical Hamiltonian vector field associated to the corresponding variable. However, ambiguities remain  when one studies the evolution across the {\em would-be-singularity} of the Kantowski-Sachs model at $a=0$ \cite{Perez:2023jrq}.   

Now we would like to concentrate on the evolution when we are away from the $a=0$.
In such regime the one step evolution \eqref{step} can be composed to produce the arbitrary initial to final area evolution 
\be
{\psi(m,{\rm p}_\phi, \epsilon+n \lambda )=
 \left( \frac{\epsilon+n \lambda}{ \epsilon+q \lambda}\right)^{\frac{im}{2} \left(1+\frac{{\rm p}_{\phi}^2}{16\pi \ell_p^2 m^2}\right)}  
\psi(m,{\rm p}_\phi, \epsilon+q \lambda)},
\ee
for arbitrary integers $n,q>1$. 

The polymer dynamics that arises from the geometric action of the quantum constraint \eqref{dydy} enjoys of the appealing feature
of being closely related to the dynamics that one would obtain in the continuum Schroedinger representation. This statement can be made precise as follows:  any solution of the Schroedinger equation
\eqref{SE} induces on any given lattice $\Gamma_{\epsilon, \lambda}$ a solution of \eqref{dydy}. Conversely, physical states of the polymer theory represent a discrete sampling of the continuum solutions of \eqref{SE}. However, the Schroedinger evolution is ill defined at the singularity $a=0$ due to the divergence of the $1/a$ factor in front of the second term of \eqref{SE}. The polymer representation allows for a well defined evolution across the singularity thanks to the deviations from the $1/a$ behaviour introduced by the analog of the `inverse-volume' corrections \cite{Ashtekar:2011ni}. 
As  discussed in \cite{Amadei:2022zwp} these corrections are ambiguous (as fact that should not be surprising given the expectation that the classical theory cannot guide us all the way to the deep UV in QFT). Instead of proposing one particular UV extension, as in the example shown in \cite{Perez:2023jrq} where Thiemann regularization was used, one might simply keep all possibilities open and assume that the corresponding operator  is regularized in the relevant region by some arbitrary function $\log(a)\to \tau(a)$. In regions where $\tau(a)=\log(a)$ the quantum evolution leads to semiclassical equations that match exactly Einstein's equations in the KS sector, when semiclassical states are considered, and the Schroedinger equation away from $a=0$.
.

\subsection{Quantum hair: the mass operator (in the vacuum case)}\label{degeM}

As mentioned at the end of Section \ref{looloo}, the huge degeneracy introduced by the presence of the $\epsilon$ sectors plays an important role in the spectrum of physical observables. In this section we will study this degeneracy for the mass operator in the vacuum case. 
The mass of the black hole (in the vacuum case) is given by the following expression in terms of the variables \ref{nsv} (see  \cite{Perez:2023jrq} for details)
\begin{equation}\label{M(m,pm)}
M=\alpha(a)m^2e^{{\rm p}_m}.
\end{equation}
where $\alpha(a)\equiv{2\sqrt{4\pi}\ell_p^4}/{(\ell_0^2\sqrt{a})}$. For the ordering $m \exp{{\rm p}_m} m$ the eigenstates are \begin{equation} \label{eigenstate M}
|M \rangle=\sum_{a \in \Gamma_{\epsilon,\lambda}} \int \phi_M({\rm p}_m,a)|{\rm p}_m\rangle| a\rangle d{\rm p}_m,
\end{equation}
where the sum runs over the discrete lattice $\Gamma_{\epsilon,\lambda}$ defined in \eqref{cocuna} when introducing the dynamical constraint \eqref{dydy}, with
\begin{equation}\label{eigen M}
\phi_M({\rm p}_m,a)\equiv \bra{{\rm p}_m,a}M \rangle=  \sqrt{\frac{2\sqrt{M}}{\alpha(a)}}e^{-\frac{{\rm p}_m}2} J_1\left(2\sqrt{\frac{M}{\alpha(a)}}e^{-\frac{{\rm p}_m}2}\right),
\end{equation}
where $J_1$ is a Bessel function. 
One can explicitly verify that the quantum dynamics \eqref{dydy} preserves the eigenstates. 
The spectrum of the mass operator is continuous. It was argued in the context of the full LQG theory in \cite{Perez:2014xca,  Amadei:2019wjp, Perez:2022jlm} that the eigenspaces of the mass should be {infinitely degenerate} due to the underlying discrete structure of the fundamental theory and the existence of defects that would not be registered in the ADM mass operator. Interestingly,  the conjectured property is illustrated explicitly in our simple toy model as the eigenvectors \eqref{eigenstate M} for a given eigenvalue $M$ there are infinitely many and labelled by a continuum parameter. More precisely they are associated with wave functions of the form \eqref{eigen M} supported on lattices with different values of $\epsilon$. Thus eigenstates of the mass should then be denoted $|M, \epsilon \rangle$ with orthogonality relation 
\be
\braket{M,\epsilon |M^\prime, \epsilon^\prime}_{\rm phys}=\delta(M,M^\prime)\delta_{\epsilon,\epsilon^\prime},
\ee  
where $\delta_{\epsilon,\epsilon^\prime}$ is the Kronecker delta symbol.
The existence of such a large degeneracy is a generic feature of the polymer (or loopy) representation. Even when this is a toy model of quantum gravity, this feature 
is likely to reflect a basic property of the representation of the algebra of observables in the full LQG context. Here we are showing that the mass operator is hugely degenerate suggesting that the usual assumption of the uniqueness of the vacuum in background dependent treatments of quantum field theory might fail  in the full quantum gravity context. The present conclusions are independent of factor ordering ambiguities.


\section{Modelization of a Hawking pair on our quantum geometry}\label{model}

Outgoing Hawking excitations detected in normalizable wave packets at future null infinity are expected to evolve unitarily---independently of the inside degrees of freedom---when they become sufficiently separated from the black hole. These excitations are however entangled with the inside falling partner. The back reaction of the inside partner on the quantum geometry is expected to be strong as the particle falls toward the singularity. Instead, the back reaction of the outside partner is expected to be weak as the evolution happens in the weak gravitational field region (the back reaction is completely neglected in the original Hawking calculation for these reasons). The setup we introduce in this section incorporates a simplified version for the back reaction for the inside partner (based on the model of the previous section). The dynamical evolution for the outside partner is simplified by assuming it to be trivial: the outside Hamiltonian vanishes. The key feature is however the entanglement in the initial state and the transfer of information to the microscopic structure of the quantum geometry which will be shown in Section \ref{resultats}.  

We have argued in the introduction why such a mechanism involves primarily the inside quantum dynamics where the strong field regime activates the interaction with the Planckian discreteness. This argument is valid in general (independent of our simplifying assumptions) and realized in particular  in our simple model. 
Accordingly, in this section we use the model introduced in the previous
section to describe the non trivial dynamics of the inside partner evolving towards the strong quantum gravity region near the singularity, while letting outside partner evolve freely with a trivial Hamiltonian.
The exact dynamics of the corresponding density matrix of the full system as well as the one corresponding to the relevant reduced density matrices of suitable subsystems will be written explicitly. 

\subsection{Construction of the initial state}

In order to simplify as much as we can the description of the matter part, we will consider that the {\em inside} matter can be in two different  orthogonal states $|+ \rangle$ or $|- \rangle \in \mathcal{\sH}_{in}$ (with subscript $in$ standing for the {\em inside} Hilbert space of the scalar degrees of freedom) which are eigenvector of $\hat {\rm p}_{\phi} $ of eigenvalue ${\rm p}_{+} $ and ${\rm p}_{-} $ respectively. On the other hand, we will consider that the outgoing part of the matter can be in two different orthogonal states $|+\rangle$ or $|- \rangle \in \mathcal{\sH}_{out}$ (with subscript $out$ standing for the {\em outside} Hilbert space of the scalar degrees of freedom). As the outside outgoing partner evolves in regions of low curvature where the type of effects we are focusing on should be very small, we model such evolution by a trivial Hamiltonian. This simple description of the matter is sufficient to write down a state for the total matter where the ingoing particle and the outgoing one are initially maximally entangled, and to study how the dynamic modifies these correlations. 
We can now write down the total Hilbert space describing the geometrical and the matter degrees of freedom. It can be written as
\begin{equation}
	\mathcal{\sH}_{total}=\underbrace{\mathcal{\sH}_{m} \otimes \mathcal{\sH}_{\epsilon}}_{\rm geometry} \otimes \overbrace{\mathcal{\sH}_{in} \otimes \mathcal{\sH}_{out}}^{\rm matter} \ ,
\end{equation}
where we have organized the Hilbert space of geometry as the tensor product of normalizable functions $\sH_m$ of the configuration variable $m$, defined in \eqref{nsv}, on a given lattice times the separable truncation $\sH_{\epsilon}=\C^{n_{\epsilon}}$ of the epsilon sector where we assume that a finite number $n_\epsilon$ of lattices $\Gamma_{\epsilon, \lambda}$---introduced in \eqref{cocuna}---are allowed. Concretely, and for simplicity, we will take $n_\epsilon=2$ with 
two alternative values $\epsilon_\pm$. With such simple choices we have that $\mathcal{\sH}_{\epsilon}= \mathcal{\sH}_{in}=\mathcal{\sH}_{out}=\C^2$,  i.e., with the exception of $\sH_m$ that is infinite dimensional, the other quantum numbers have been truncated to standard $q$-bits.  Thus,  
a basis of this Hilbert space is defined by the set of vectors $| m,\varepsilon,i,o \rangle$ with $m \in \mathbb{R}^+$, $\varepsilon$ (encoding the microscopic $\epsilon$ sectors), $i$ (encoding the {\em inside} particle state) and $o$ (encoding the {\em outside} state) spin-like quantum numbers taking values $\pm$. The inner product between basis states is 
\begin{equation}
	\langle m_1,\varepsilon_1,i_1,o_1 | m_2, \varepsilon_2, i_2 ,o_2 \rangle =\delta(m_1-m_2) \delta_{\varepsilon_1,\varepsilon_2}\delta_{i_1,i_2}\delta_{o_1,o_2} \ .
\end{equation}
After having defined the truncation of the Hilbert space we introduce the initial state representing a toy version of the in-vacuum in the context of Hawking radiation, namely 
\begin{equation} \label{initial state}
	|\psi_0 \rangle =  \sum_{\substack{i=\pm \\ \varepsilon=\pm}}\int dm\ \psi_0(m,a_0+\epsilon_\varepsilon)|m,\varepsilon,i,i\rangle  ,  
\end{equation}
and the normalization condition $\langle \psi_0 | \psi_0 \rangle = 1$ implying
\begin{equation} \label{norm}
	\int dm\ |\psi_0(m,a_0+\epsilon_\varepsilon)|^2 =\frac{1}{4} \ .
\end{equation}
The key feature of the initial state is that it represents a state where the excitations in $\sH_{in}$ and $\sH_{out}$ are maximally correlated 
as in the standard in-vacuum.
We can then apply the evolution law \eqref{dydy} to this state, a shift by $n\lambda$ is given by
\begin{align}\label{final state}
	|\psi \rangle &=  \sum_{\substack{i=\pm \\ \varepsilon=\pm}}\int  dm\ e^{\frac{i}{2}\log\left(1+\frac{n\lambda} {a_0+\epsilon_\varepsilon}\right)\left(m+\frac{{\rm p}^2_{i}}{16\pi \ell_p^2m}\right)}\psi_0(m,a_0+\epsilon_\varepsilon)|m,\epsilon_\varepsilon,i,i\rangle \\
	&= \sum_{\substack{i=\pm \\ \varepsilon=\pm}}\int dm \  C(m,\epsilon_\varepsilon,i,n)\ |m,\epsilon_\varepsilon,i,i\rangle \ ,
\end{align}
where, in order to simplify the notation, we introduced
\begin{equation} \label{def C}
	C(m,\epsilon_\varepsilon,i,n)=e^{\frac{i}{2}\log\left(1+\frac{n\lambda} {a_0+\epsilon_\varepsilon}\right)\left(m+\frac{{\rm p}^2_{i}}{16\pi \ell_p^2m}\right)}\psi_0(m,a_0+\epsilon_\varepsilon) \ .
\end{equation}
The previous is the quantum gravitational dynamics of the initial state (of maximally correlated Hawking partners). This dynamics encodes the back reaction of the Hawking particles with the quantum geometry exactly (in the sense of our simplified model). It only remains to study the dynamical evolution of the entanglement between the relevant degrees of freedom as the {\em inside} partner evolves toward the strong quantum gravity regime (the singularity) inside of the black hole.

\subsection{The subsystems of interest and their reduced density matrices}
We now have access to the density matrix of the system $\rho=|\psi\rangle \langle \psi|$  which is a pure state at any area $a$-time (by the unitarity of the quantum evolution \ref{dydy}). Its expression is given by 

\begin{equation} \label{dm}
	\rho=\sum_{\substack{i,j=\pm \\ \varepsilon,\varepsilon'=\pm }}\int dmdm'\ C(m,\epsilon_\varepsilon,i,n) \overline{C(m',\epsilon_{\varepsilon'},j,n)}|m,\epsilon_\varepsilon,i,i\rangle \langle m',\epsilon_{\varepsilon'},j,j|  .
\end{equation}
We determine now the reduced density matrices of the relevant subsystems. We will start by writing the density matrix of the subsystem $\mathcal{\sH}_{\epsilon}\otimes\mathcal{\sH}_{in}$ obtained from tracing out the 
$\sH_m$ and $\sH_{out}$. One can show that 
\begin{equation} \label{dm ei}
	\rho^{\epsilon,\textrm{in}}=
	\begin{pmatrix}
		1/4 & 0 & D(\epsilon_1,\epsilon_2,+,+,n) & 0\\
		0 & 1/4 & 0 & D(\epsilon_1,\epsilon_2,-,-,n)\\
		\overline{D(\epsilon_1,\epsilon_2,+,+,n)} & 0 & 1/4 & 0\\
		0 & \overline{D(\epsilon_1,\epsilon_2,-,-,n)} & 0 & 1/4
	\end{pmatrix} \ ,
\end{equation}
where \begin{equation} \label{def D}
	D(\epsilon_\varepsilon,\epsilon_{\varepsilon'},i,j; n)\equiv \int dm \ C(m,\epsilon_\varepsilon,i,n)\overline{C(m,\epsilon_{\varepsilon'},j,n)} , 
\end{equation}
Tracing further over $\sH_{in}$ we get the reduced density matrix of the subsystem $\mathcal{\sH}_{\epsilon}$ 
\begin{equation} \label{dm e}
	\rho^{\epsilon}=
	\begin{pmatrix}
		1/2 & D(\epsilon_1,\epsilon_2,+,+,n)+D(\epsilon_1,\epsilon_2,-,-,n)   \\
		\overline{D(\epsilon_1,\epsilon_2,+,+,n)} +\overline{D(\epsilon_1,\epsilon_2,-,-,n)}  &1/2 
	\end{pmatrix} \ ,
\end{equation}
while tracing over $\sH_{\epsilon}$ we obtain the reduced density matrix of the subsystem ${\mathcal{\sH}_{in}}$  
\begin{equation} \label{dm i}
	\rho^{\textrm{in}}=
	\begin{pmatrix}
		1/2 & 0  \\
		0 & 1/2 
	\end{pmatrix} \ .
\end{equation}
One can notice that this last density matrix does not depend on time, which means that the entanglement entropy of this subsystem is constant. More generally, the results \eqref{dm ei}, \eqref{dm e} and \eqref{dm i} will be used in the next section to study the correlations between the two subsystems ${\mathcal{\sH}_{\epsilon}}$ and ${\mathcal{\sH}_{in}}$. But we are also interested by the evolution of the correlations between the subsystems ${\mathcal{\sH}_{in}}$ and ${\sH_{out}}$. Consequently, we need the reduced density matrix associated to the subsystem $\sH_{in}\otimes\sH_{out}$. 
It follows that \begin{equation} \label{dm io}
	\rho^{\textrm{in},\textrm{out}}=
	\begin{pmatrix}
		1/2 & 0  & 0 & D(\epsilon_1,\epsilon_1,-,+,n)+D(\epsilon_2,\epsilon_2,-,+,n)\\
		0 & 0 & 0 & 0\\
		0 & 0 & 0 & 0\\
		\overline{D(\epsilon_1,\epsilon_1,-,+,n)+D(\epsilon_2,\epsilon_2,-,+,n)} & 0 & 0 & 1/2
	\end{pmatrix} \ .
\end{equation}
Finally, the last ingredient that one will need is the matrix density of the subsystem $\sH_{out}$ given by
\begin{equation} \label{dm o}
	\rho^{\textrm{out}}=
	\begin{pmatrix}
		1/2 & 0  \\
		0 & 1/2 
	\end{pmatrix} \ .
\end{equation}

\subsection{Computation of the interference terms}

From the previous section, we have a complete knowledge for any $a$ of the total system, cf \eqref{dm}, and also of the relevant subsystems $\sH_{in}$ (matter inside), $\sH_{out}$ (matter outside), $\sH_{\epsilon}$ (quantum hair), $\sH_{in}\otimes\sH_{\epsilon}$, and $\sH_{in}\otimes\sH_{out}$ respectively given by \eqref{dm i}, \eqref{dm o}, \eqref{dm e}, \eqref{dm ei} and \eqref{dm io}. We now want to study the entanglement of these subsystems with their environment which is encoded in the non diagonal terms of the different density matrices characterized by the complex function  $D(\epsilon, \epsilon', i, j; n)$ appearing in \eqref{dm ei}, \eqref{dm e} and \eqref{dm io}. In order to evaluate  $D(\epsilon, \epsilon', i, j; n)$, we need to choose the form of initial wave function $\psi_0(m,a_0,{\rm p}_i)$ of the quantum geometry in \eqref{initial state}. We take the following semiclassical state
\begin{equation} \label{initial gauss}
	\psi_0(m,a_0+\epsilon_\varepsilon) = \frac{1}{2\sqrt{\sigma\sqrt{\pi}}}e^{-\frac{(m-m_0)^2}{2\sigma^2}}, 
\end{equation}
normalized according to  \eqref{norm}, and centered at $m=m_0$ (with a spread $2\sigma$) and ${\rm p}_m=0$. Moreover, we assume that $m_0 \gg  \sigma $.
According to \eqref{M(m,pm)}---giving initial conditions at $a=100\lambda \ell_p^2$ (an arbitrary choice of initial time)---such state corresponds to a semiclassical state of the spacetime geometry corresponding to a Schwarzschild black hole of mass \be \label{massa}M=\frac{{\sqrt{4\pi}\ell_p^3 m_0^2}}{5 \ell_0^2\sqrt{\lambda}}.\ee The relevant 
amplitude $D(\epsilon, \epsilon', i, j)$ becomes
\begin{equation} \label{def C 2}
	D(\epsilon_\varepsilon,\epsilon_{\varepsilon'},i,j; n)= \frac{1}{4\sigma\sqrt{\pi}}\int_{-\infty}^{+\infty} dm \  e^{-\frac{(m-m_0)^2}{\sigma^2}+\frac{i}{2}\log\left(1+\frac{n\lambda} {a_0+\epsilon_\varepsilon}\right)\left(m+\frac{{\rm p}^2_{i}}{16\pi \ell_p^2m}\right)-\frac{i}{2}\log\left(1+\frac{n\lambda} {a_0+\epsilon_{\varepsilon'}}\right)\left(m+\frac{{\rm p}_j^2}{16\pi \ell_p^2m}\right)}.
\end{equation}
%
In order to compute the previous integral we first change variables to $k=m-m_0$, with $k \in [-\infty,+\infty]$ and approximate the integral using the stationary phase method (Gaussian integrals). The result is 
\begin{align}\label{approx D}
	D(\epsilon_\varepsilon,\epsilon_{\varepsilon'},i,j; n)\approx&\frac{e^{\frac{1}{2} i \left(\frac{{\rm p}^2_{i}}{16 \pi  \ell_p^2 m_0}+m_0\right) \log \left(\frac{a_0+\lambda n+\epsilon_\varepsilon}{a_0+\epsilon_\varepsilon}\right)-\frac{1}{2} i \left(\frac{{\rm p}_j^2}{16 \pi  \ell_p^2 m_0}+m_0\right) \log \left(1+\frac{n\lambda}{a_0+\epsilon_{\varepsilon'}}\right)}}{4\sigma\sqrt{\pi}}
	\notag
	\\
	&\times
	\int_{-\infty}^{+\infty} e^{k^2 \left(\frac{i {\rm p}^2_{i} \log \left(1+\frac{n\lambda}{a_0+\epsilon_\varepsilon}\right)}{32 \pi  \ell_p^2 m_0^3}-\frac{i {\rm p}_j^2 \log \left(1+\frac{n\lambda}{a_0+\epsilon_{\varepsilon'}}\right)}{32 \pi  \ell_p^2 m_0^3}-\frac{1}{\sigma ^2}\right)}
	\notag
	\\
	&\times		
	e^{k \left(\frac{1}{2} i \left(1-\frac{{\rm p}^2_{i}}{16 \pi  \ell_p^2 m_0^2}\right) \log \left(1+\frac{n\lambda}{a_0+\epsilon_\varepsilon}\right)-\frac{1}{2} i \left(1-\frac{{\rm p}_j^2}{16 \pi  \ell_p^2 m_0^2}\right) \log \left(\frac{a_0+\lambda  n+ \epsilon_{\varepsilon'}}{a_0+\epsilon_{\varepsilon'}}\right)\right)} dk .
\end{align}

This integral is analytically computable and is equal to

\begin{align}\label{final D}
	D(\epsilon_\varepsilon,\epsilon_{\varepsilon'},i,j; n)\approx& \frac 1{\sigma  \sqrt{-\frac{i {\rm p}^2_{i} \log \left(1+\frac{n\lambda}{a_0+\epsilon_\varepsilon}\right)}{\ell_p^2 m_0^3}+\frac{i {\rm p}_j^2 \log \left(1+\frac{n\lambda}{a_0+\epsilon_{\varepsilon'}}\right)}{\ell_p^2 m_0^3}+\frac{32 \pi }{\sigma ^2}}}
\notag
	\\
	&\times
	\exp \left(-\frac{\sigma ^2 \left(\left({\rm p}^2_{i}-16 \pi  \ell_p^2 m_0^2\right) \log \left(1+\frac{n\lambda}{a_0+\epsilon_\varepsilon}\right)-\left({\rm p}_j^2-16 \pi  \ell_p^2 m_0^2\right) \log \left(1+\frac{n\lambda}{a_0+\epsilon_{\varepsilon'}}\right)\right)^2}{128 \pi  \ell_p^2 m_0 \left(-i {\rm p}^2_{i} \sigma ^2 \log \left(1+\frac{n\lambda}{a_0+\epsilon_\varepsilon}\right)+i {\rm p}_j^2 \sigma ^2 \log \left(1+\frac{n\lambda}{a_0+\epsilon_{\varepsilon'}}\right)+32 \pi  \ell_p^2 m_0^3\right)}\right)
	\notag
	\\
	&\times			
	{\sqrt{2 \pi } \left(1+ \frac{\lambda  n}{a_0+\epsilon_\varepsilon}\right)^{\frac{1}{2} i \left(\frac{{\rm p}^2_{i}}{16 \pi  \ell_p^2 m_0}+m_0\right)} \left(1+\frac{\lambda  n
	}{a_0+\epsilon_{\varepsilon'}}\right)^{-\frac{1}{2} i \left(\frac{{\rm p}_j^2}{16 \pi  \ell_p^2 m_0}+m_0\right)} } .\end{align}

\section{Evolution of the entanglement and correlations between the different subsystems}\label{resultats}

We have all we need to analyze the evolution of entanglement between the relevant degrees of freedom as the system evolves toward the singularity at $a=0$.
The question that remains is what measure of entanglement we use for interpretation. Here there are several alternatives which all encode the same 
physics in different terms. The simplest (yet to some extend basis dependent method) is the use of the so-called decoherence function. Mutual information is basis independent and provides a more clear ground for interpretation and will be defined in Section \ref{mumu}.

\subsection{Decoherence function}

\begin{figure}[h!!!!!!!!!]
	\centering
	\subfigure[]
	{\includegraphics[scale=0.2]{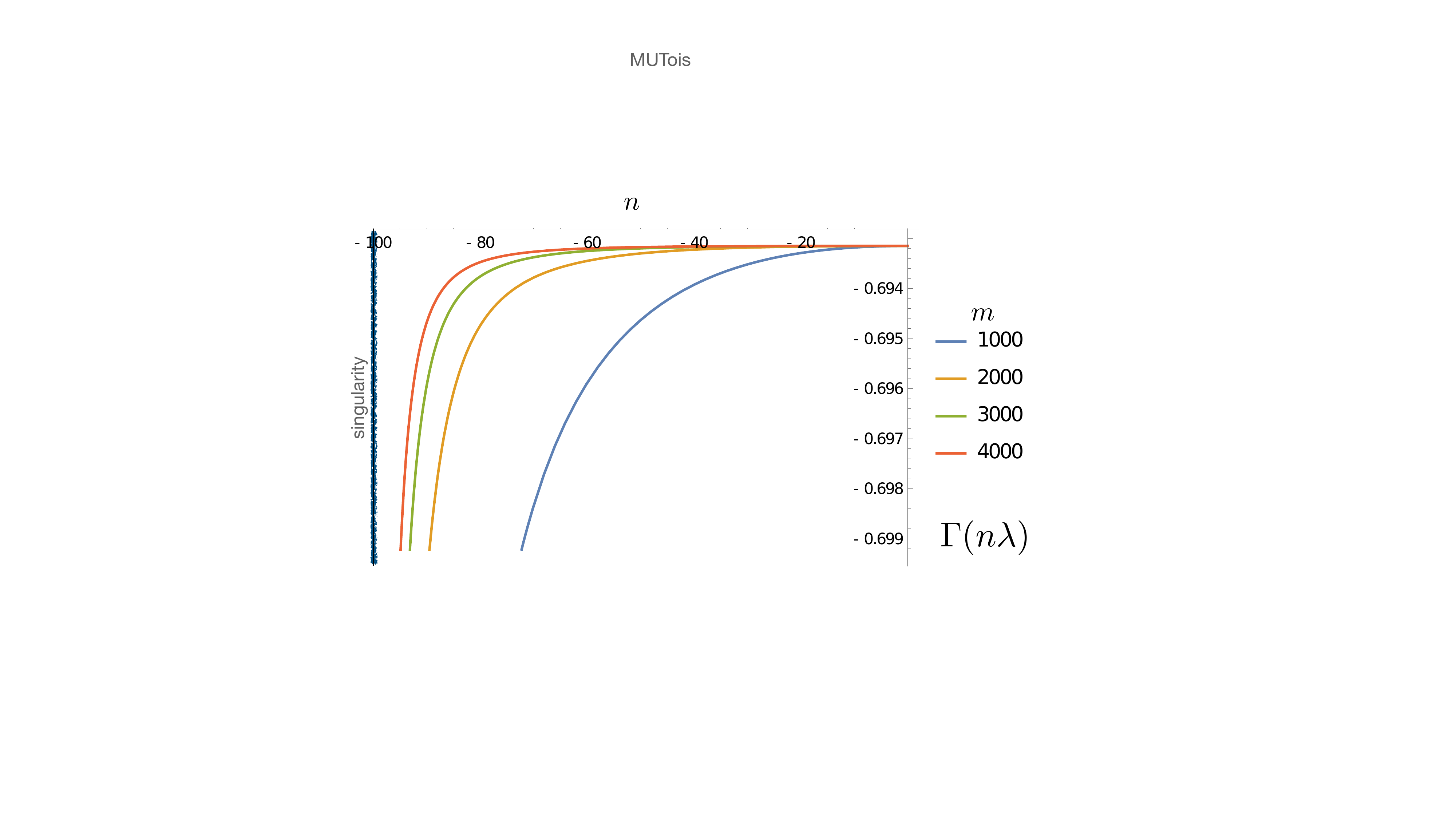}}
	\hspace{1mm}
	\subfigure[]
	{\includegraphics[scale=0.2]{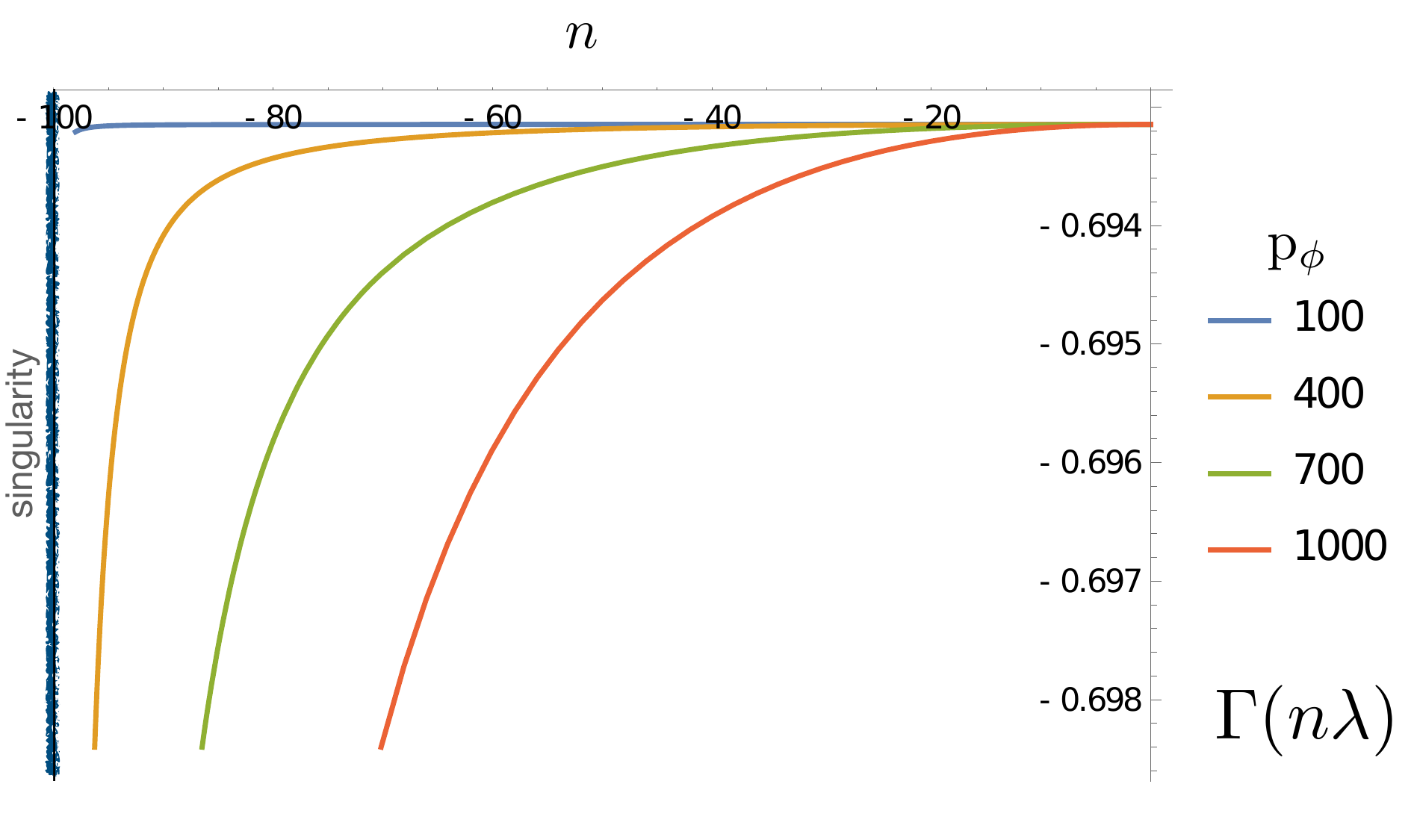}}
	\hspace{1mm}
	\subfigure[]
	{\includegraphics[scale=0.2]{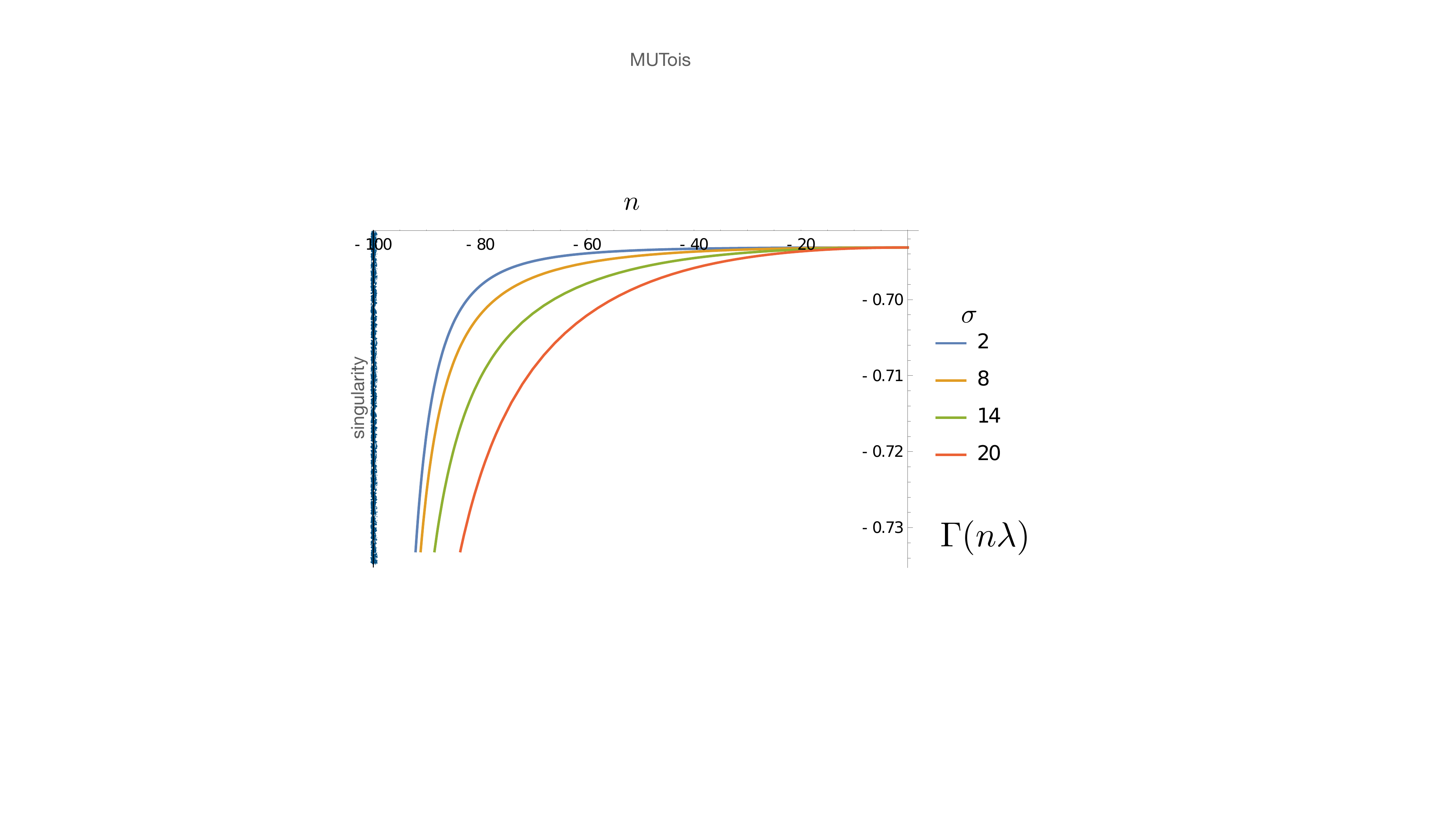}}
	\caption{(a) Plot of $\Gamma$ in terms of $n$ evolution steps  for $\sigma=10$, $p_+=1000$ and for different values of $m_0$. (b) Plot of $\Gamma$ in terms of $n$ evolution steps  for $m_0=1000$, $\sigma=10$ and for different values of $p_+$. (c) Plot of $\Gamma$ for $m_0=1000$, $p_+=1000$ and for different values of $\sigma$. The singularity correspond to the value $n\lambda=-100$ and is indicated by a black vertical line.}
	\label{fig:decoh}
\end{figure}

It is often said that the entanglement of a system with its environment tend to kill the off diagonal terms of its density matrix. The consequence is that a system initially pure becomes mixed due to this interaction. It is possible to make a more precise statement thanks to the decoherence functions. Let us consider a general subsystem $\rho(t)$ written in a basis that we will note $\{ \ket{n}\}_{n \in \mathbb{N}}$. This subsystem is initially pure and interacts with its environment. The decoherence functions will be noted $\Gamma_{n m}(t)$ and are define in the following way

\begin{equation}
|\bra{n}\rho (t) \ket{m}|=\exp(\Gamma_{n m}(t)).
\end{equation} 
These functions describe the time evolution of the off diagonal terms of the density matrix and therefore encode the evolution of a subsystem from a pure state to a mixed one. For a system of dimension N there are ${N(N-1)}/{2}$ independent off diagonal terms in the density matrix, and as many decoherence functions. One can notice that these functions depend on the basis we chose to write the density matrix. There is however a natural basis, the `energy basis',  in which the diagonal terms are time independent . We will work in this basis here.
We are interested first in the evolution of the purity of the $\epsilon$ subsystem (i.e. the Planckian geometrical degrees of freedom). One can see that it is easy to obtain the decoherence function for this subsystem, which is unique, from the expression \eqref{dm e}

\begin{equation}
\Gamma(n\lambda)=|D(\epsilon_1,\epsilon_1,-,+,n)+D(\epsilon_2,\epsilon_2,-,+,n)|.	
\end{equation}
One can now plot the evolution of this function in terms of the time variable $n\lambda$ in order to illustrate how the purity of the subsystem $\epsilon$ evolves, and how this evolution depends of the different parameter of the system $m_0,p_+,\sigma$ (cf Fig \ref{fig:decoh}).

\subsection{Mutual information as an entanglement measure}\label{mumu}

The previous was a quick illustration about how the microscopic states actually entangle during the dynamical evolution and hence cannot be neglected in discussions of unitarity. However, the method used is not adapted for a clear interpretation due to its intrinsic basis dependent nature. Here we interpret the dynamical evolution of entanglement using a basis independent  measure of entanglement: the mutual information. 
We first briefly remind the reader of a few definitions and then apply them to our system.  The first ingredient we need is the notion of relative entropy. Relative entropy is a entanglement measure defined between two states described by the density matrix $\rho$ and $\sigma$ in the following way 
\begin{equation} \label{relat ent}
S(\rho | \sigma) : = {\rm Tr}(\rho \log \rho - \rho \log \sigma).
\end{equation}
One can show that $S(\rho | \sigma) \ge 0$ and it is equal to zero when $\rho=\sigma$. It quantifies the distinguishability of $\rho$ from $\sigma$. An important property of the relative entropy is that for any bounded operator $\hat O$, the relative entropy satisfies
\begin{equation} \label{prop rel ent}
\frac{\left({\rm Tr}(\hat O \rho)-{\rm Tr}(\hat O \sigma)\right)^2}{||\hat O||} \le S(\rho | \sigma),
\end{equation}
with $||\hat O || =\text{sup}\sqrt{{\bra{\psi}\hat O \ket{\psi}}/{\bra{\psi}\ket{\psi}}}$. The mutual information is a specific case of the relative entropy. Let us consider a system which can be decomposed into three subsystem $A$, $B$ and $C$. We assume that the total system is in a pure state $\ket{\psi}$, the subsystem $A+B$ is described by the density matrix $\rho_{AB}$ and the two subsystem $A$ and $B$ are respectively described by the density matrix $\rho_A$ and $\rho_B$. The mutual information (between $A$ and $B$ when $C$ is traced out) is given by
\begin{equation}
I_{AB|C} := S(\rho_{AB} | \rho_{A}\otimes\rho_{B}) .  
\end{equation}
The mutual information is therefore also an entanglement measure and quantifies how much a mixed states is distinguishable from the uncorrelated state that we can get by treating the different subsystems as independent. By using $\eqref{relat ent}$, one can show that 
\begin{equation}
I_{AB|C} = S_A + S_B - S_{AB},
\end{equation}
where $S$ is the usual Von Neumann entropy. An important property of the mutual information comes from $\eqref{prop rel ent}$, which, in the case $\rho =\rho_{AB}$ and $\sigma = \rho_{A}\otimes\rho_{B}$ gives, for all $\hat O =\hat O_A \otimes \hat O_B$ such that $\hat O_A$ and $\hat O_B$ are bounded, the following relation
\begin{equation}
	\frac{\left(\bra{\psi}\hat O_A\hat O_B \ket{\psi}-\bra{\psi}\hat O_A\ket{\psi}\bra{\psi}\hat O_B \ket{\psi}\right)^2}{2||\hat O_A||||\hat O_B||}\le I_{AB | C}(\ket{\psi}) . 
\end{equation} 
This relation means that the mutual information is an upper bound of the correlations between the subsystems $A$ and $B$. Thus, we will use this entanglement measure to track the correlations between the different subsystems in our model because  it has a clear physical meaning in terms of entanglement and correlations.

\subsubsection{Evolution of the correlations between the Hawking partners}

\begin{figure}[h!]
	\centering
	\subfigure[]
	{\includegraphics[scale=0.2]{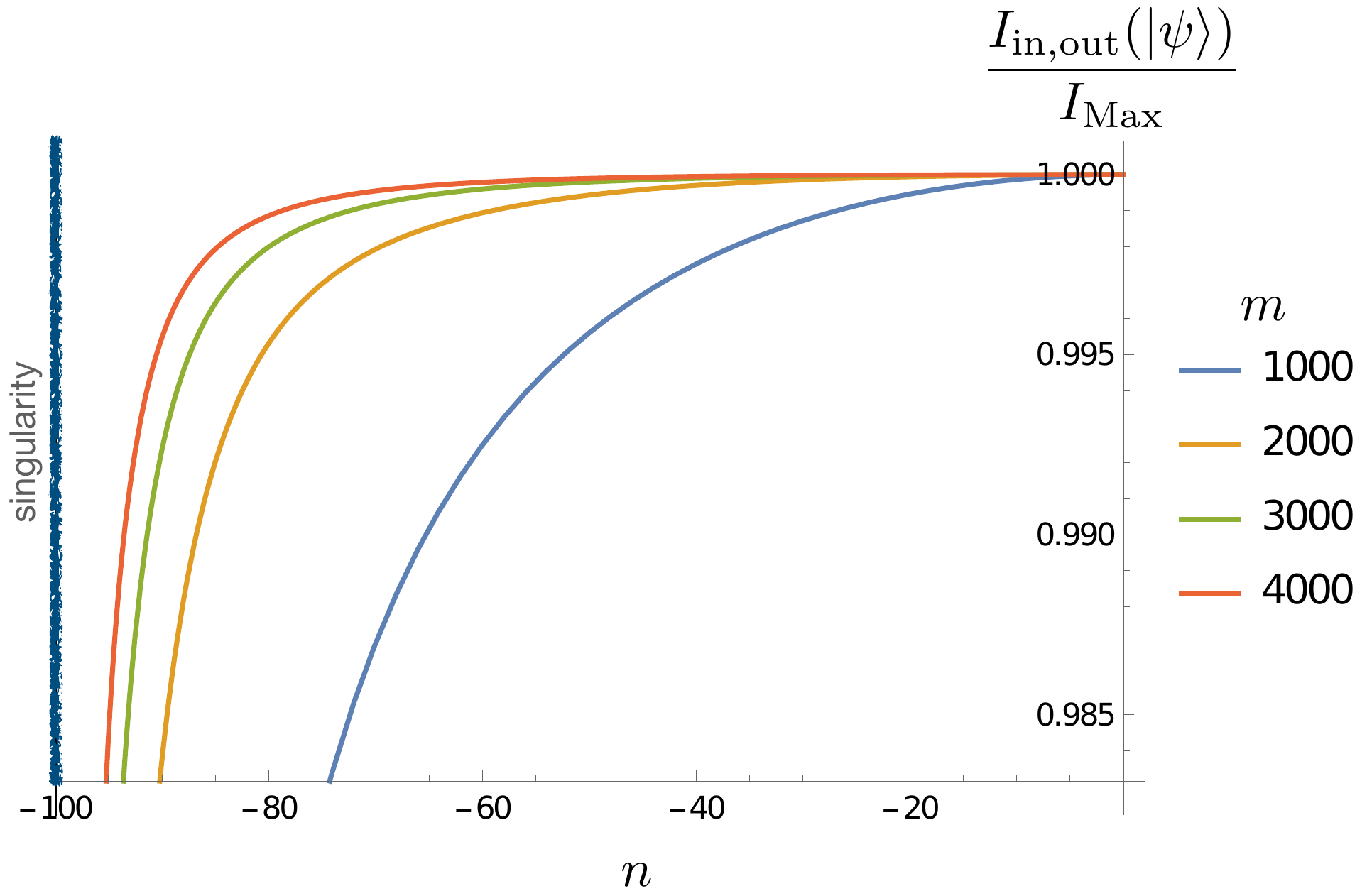}}
	\hspace{1mm}
	\subfigure[]
	{\includegraphics[scale=0.2]{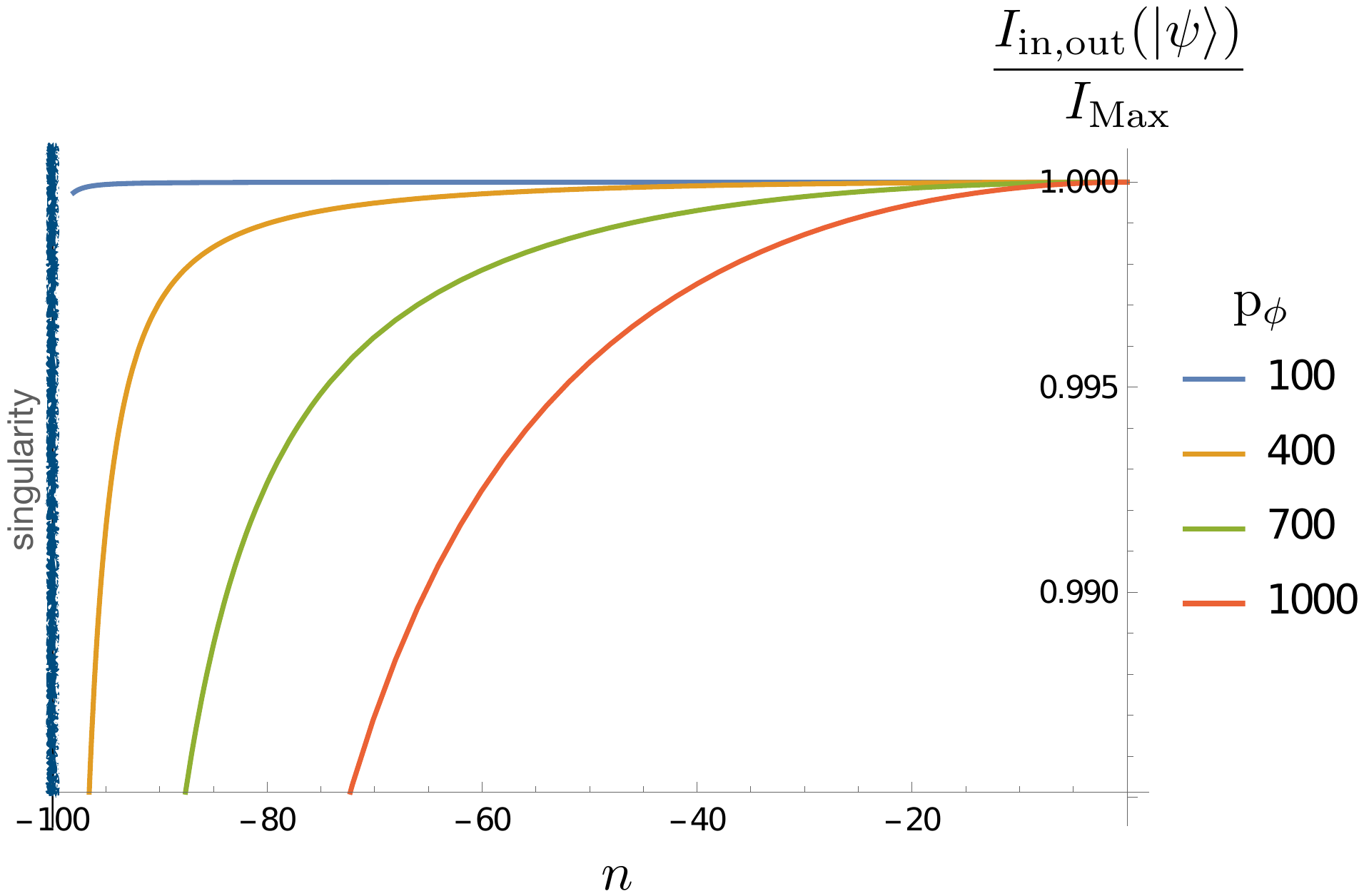}}
	\hspace{1mm}
	\subfigure[]
	{\includegraphics[scale=0.2]{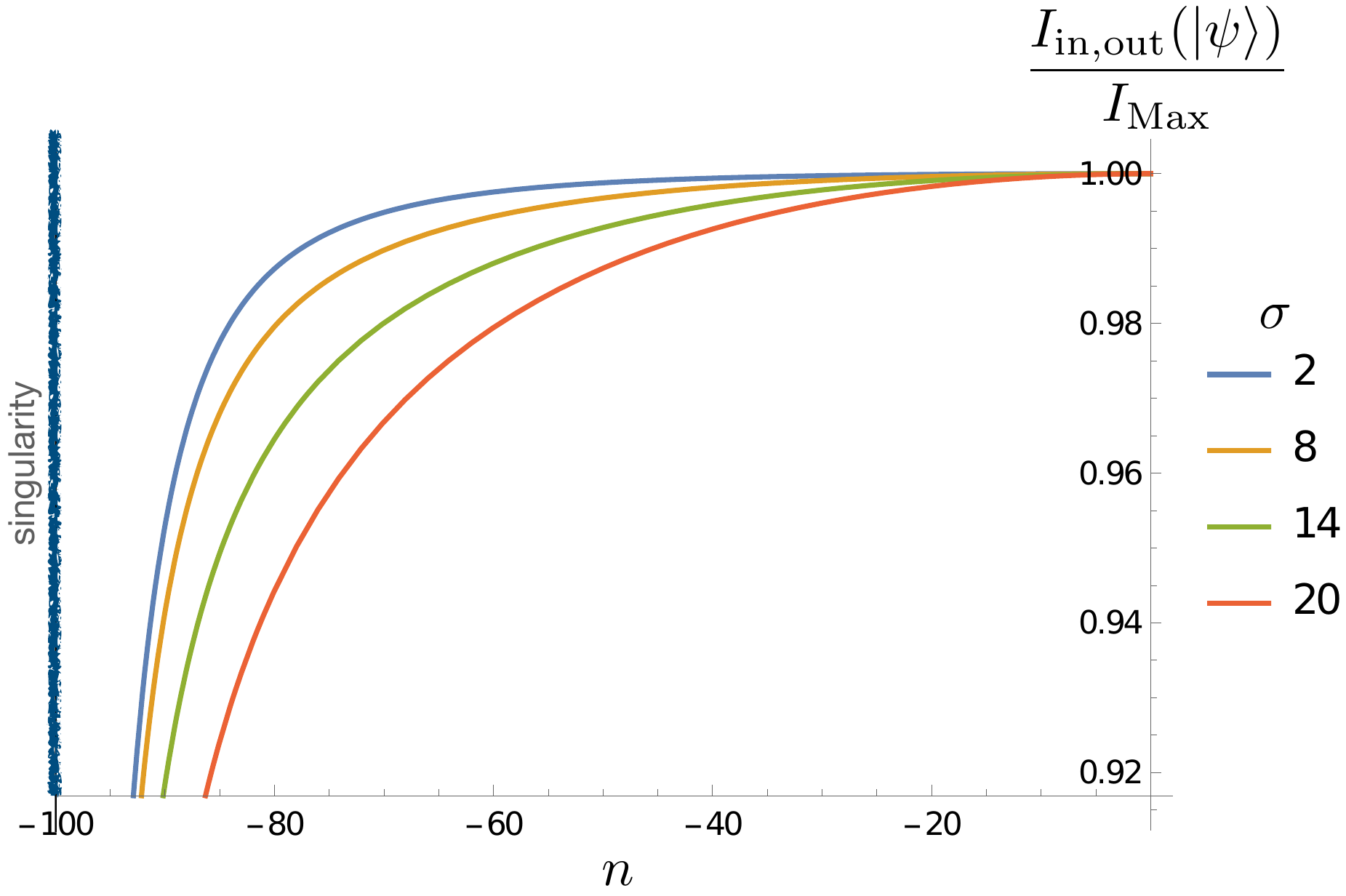}}
	\caption{(a) Plot of ${I_{\textrm{in},\textrm{out}}(|\psi\rangle)}/{I_{\text{Max}}}$ in term of $n$ evolution steps for $\sigma=10$, $p_+=1000$ and for different values of $m_0$. (b) Plot of ${I_{\textrm{in},\textrm{out}}(|\psi\rangle)}/{I_{\text{Max}}}$ in terms of $n$ evolution steps  for $m_0=1000$, $\sigma=10$ and for different values of $p_+$. (c) Plot of ${I_{\textrm{in},\textrm{out}}(|\psi\rangle)}/{I_{\text{Max}}}$ in terms of $n$ evolution steps  for $m_0=1000$, $p_+=1000$ and for different values of $\sigma$. The singularity correspond to the value $n\lambda=-100$ and is indicated by a black vertical line.}
	\label{fig:Figure2}
\end{figure}

We can compute the mutual information between the inside and the outside Hawking excitations in the matter sector
\begin{equation}\label{I in out}
	I_{\textrm{in},\textrm{out}}(|\psi\rangle)=S(\rho^{\textrm{in}})+S(\rho^{\textrm{out}})-S(\rho^{\textrm{in},\textrm{out}})
\end{equation}
From  \eqref{dm i},\eqref{dm o} and \eqref{dm io}  it follows that
\begin{align}
	I_{\textrm{in},\textrm{out}}(|\psi\rangle)&=\frac{1}{2}\left(1-2|D(\epsilon_1,\epsilon_1,j,0,n)+D(\epsilon_2,\epsilon_2,j,0,n)|\right)\log\left(\frac{1}{2}\left(1-2|D(\epsilon_1,\epsilon_1,j,0,n)+D(\epsilon_2,\epsilon_2,j,0,n)|\right)\right)
	\notag
	\\
	&+\frac{1}{2}\left(1+2|D(\epsilon_1,\epsilon_1,j,0,n)+D(\epsilon_2,\epsilon_2,j,0,n)|\right)\log\left(\frac{1}{2}\left(1+2|D(\epsilon_1,\epsilon_1,j,0,n)+D(\epsilon_2,\epsilon_2,j,0,n)|\right)\right) 
	\notag
	\\
	&+\log(4),
\end{align}
which is the final explicit result if we use \eqref{final D}. We can now look at the evolution of the mutual information between the inside and outside Hawking pairs as a function of $a=n\lambda\ell_p^2$, for $n\lambda \ell_p^2 \in [0,a_0-2\lambda\ell_p^2]$. In other words, we can visualize the evolution of the entanglement/correlations between these two subsystems when the area decreases. As we can anticipate, this quantity will depend on the choice of $m_0$ (recall that the mass $M\propto m_0^2$ from \eqref{massa}), on the spread of the wave packet $\sigma$, and on the square of the `characteristic energy' of the Hawking particle $j$.

The results are shown in Figure \ref{fig:Figure2}, where we observe the decrease of the mutual information between the Hawking partners (initially maximally correlated as in Hawking radiation) as the inside partner evolves toward the singularity. This is an expected feature: due to the back reaction of the inside particle on the quantum geometry correlations are transferred to degrees of freedom other than those encoded in $\sH_{in}$. This tell us that an appropriate account of unitarity needs the inclusion of all the degrees of freedom. From such general perspective the result is expected; a trivial implication of the monogamy of entanglement in quantum mechanics. The non trivial feature here, that will become clear in Section \ref{I m e}, is that essential correlations are established with the microscopic degrees of freedom (modelled here by the $\epsilon$ sectors). 

Figure \ref{fig:Figure2} shows that the decoherence effect (the decrease of the mutual information as the inside partner approaches the singularity) is `faster' when the black hole is smaller (when $m_0$ decreases). The effect is less important initially for macroscopic black holes even when it always becomes relevant when the singularity approaches. Larger ${\rm p}_{\phi}$ increases the effect as expected from the fact that this increases the back reaction on the geometry and microscopic sectors. The dependence on the spread $\sigma$ is consistent with the present analysis.

\subsubsection{Evolution of the correlations between the in falling matter and the $\epsilon$ d.o.f } \label{I m e}

\begin{figure}[h!]
	\centering
	\subfigure[]
	{\includegraphics[scale=0.2]{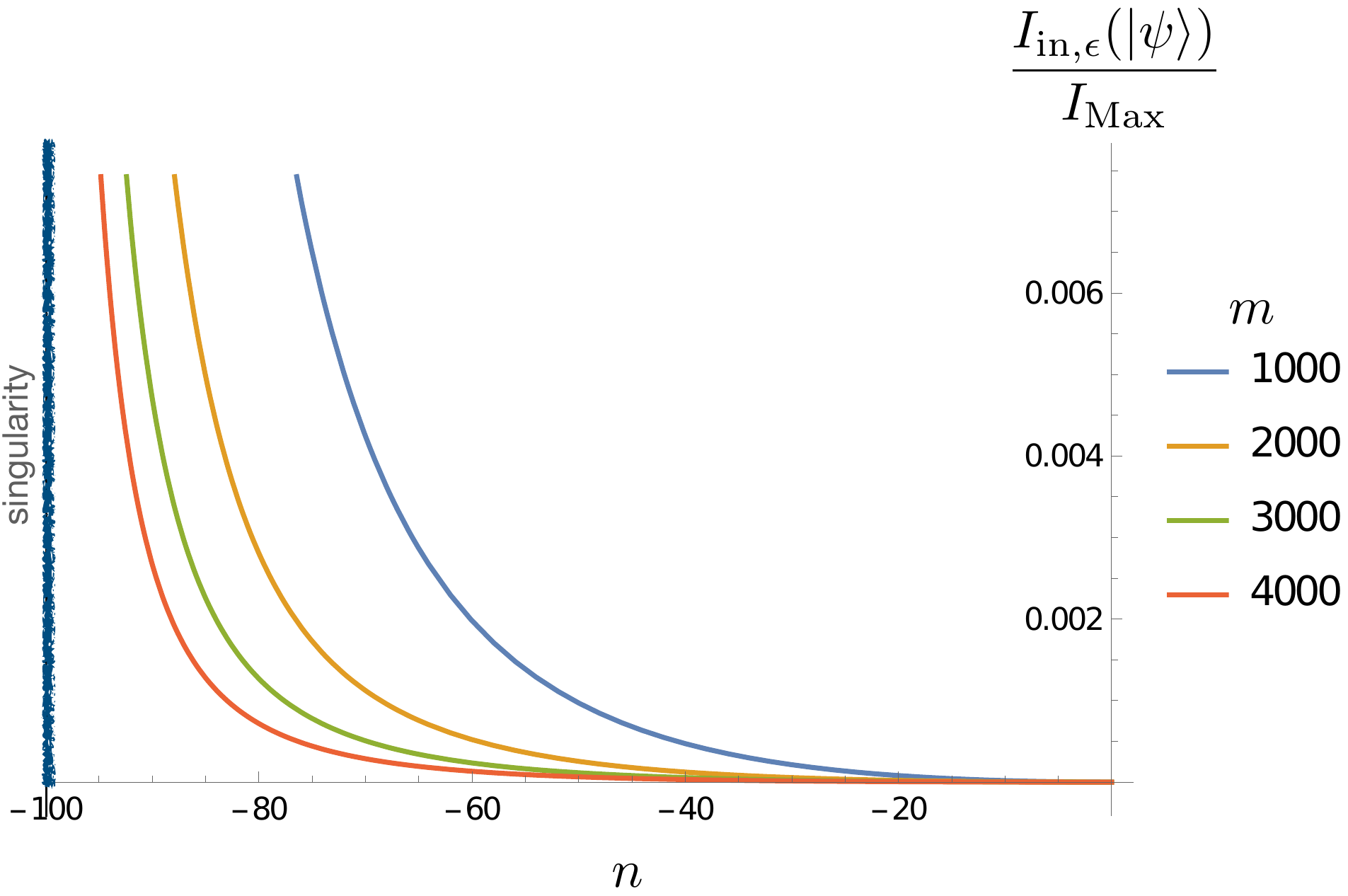}}
	\hspace{1mm}
	\subfigure[]
	{\includegraphics[scale=0.2]{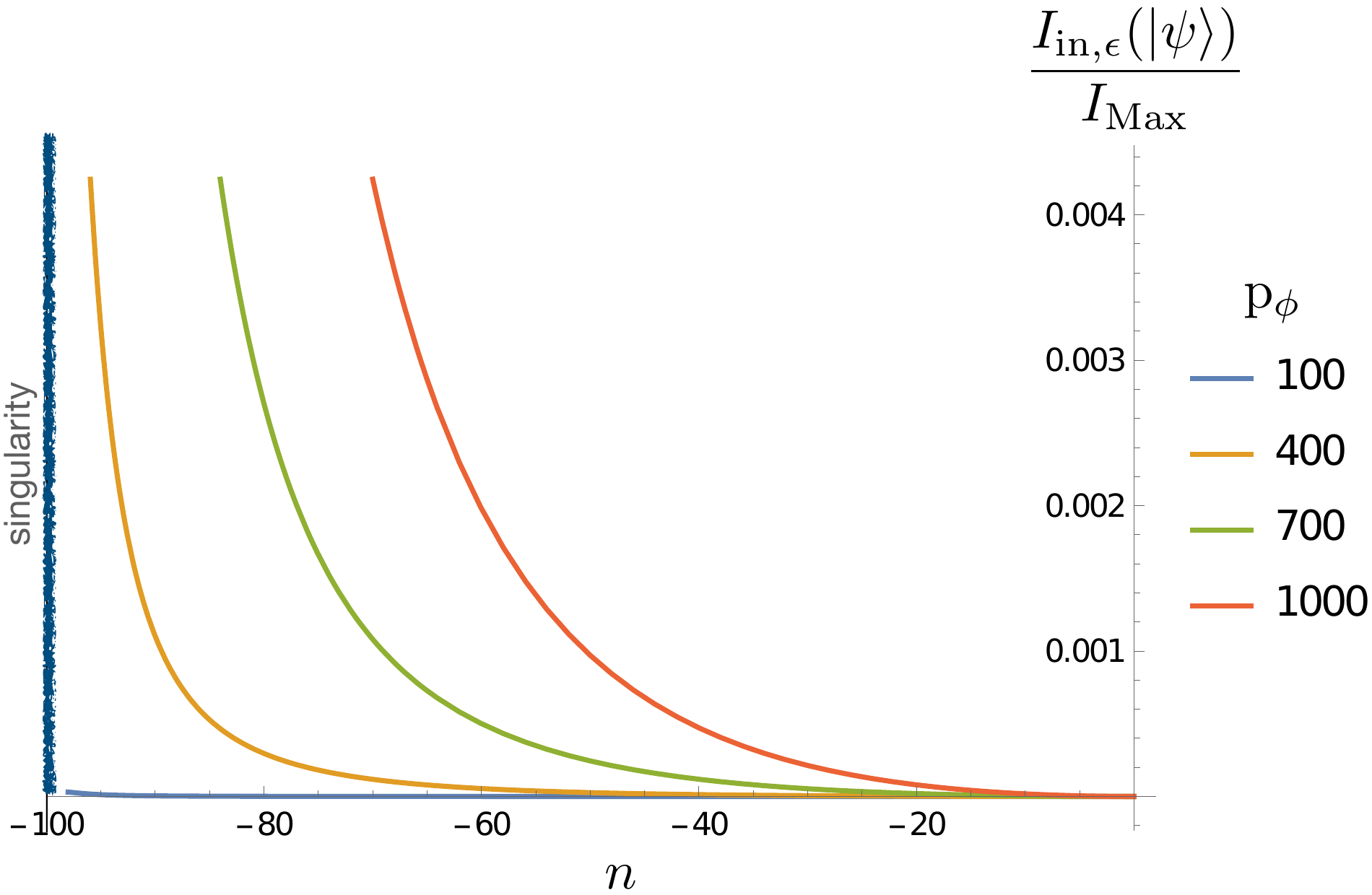}}
	\hspace{1mm}
	\subfigure[]
	{\includegraphics[scale=0.2]{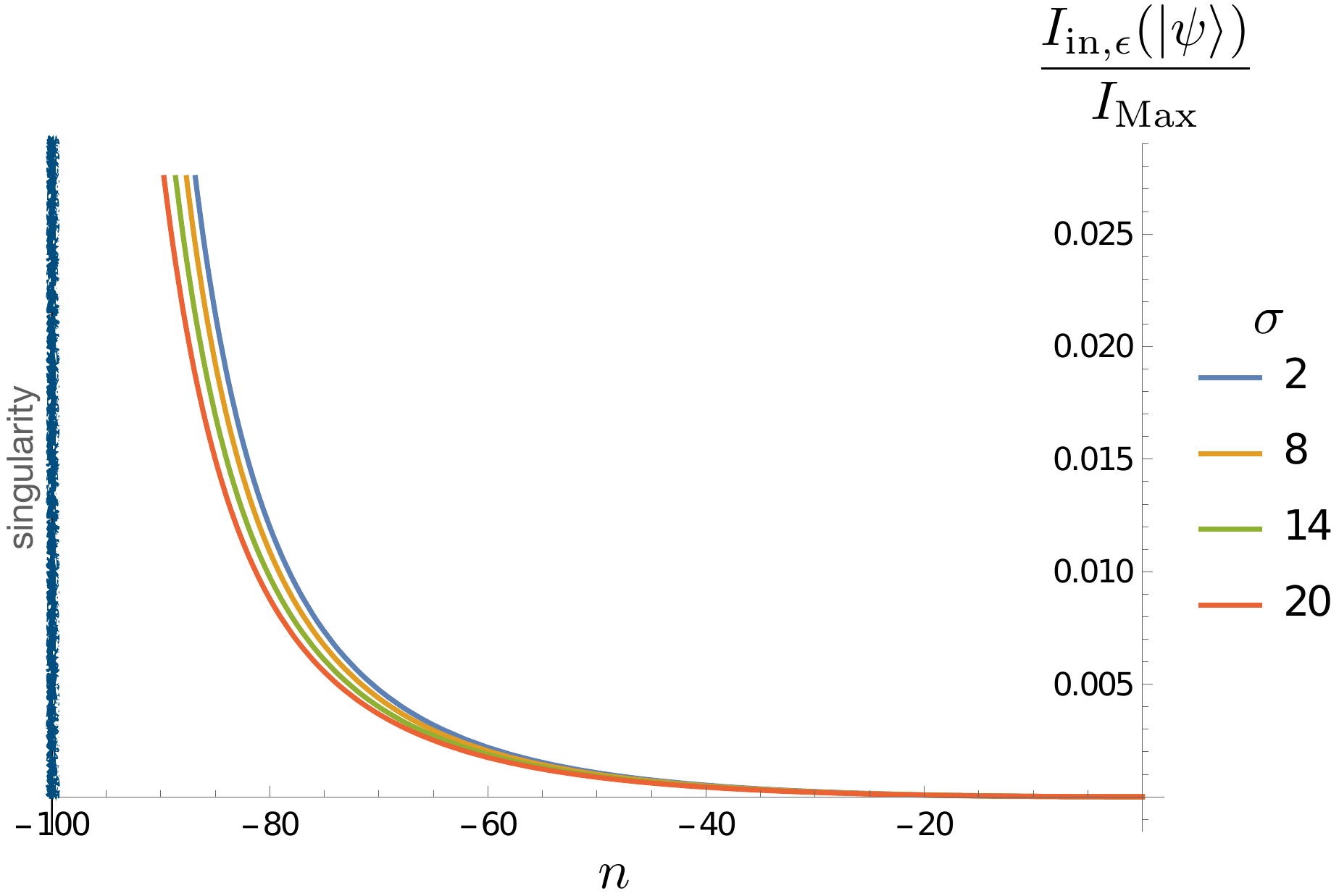}}
	\caption{(a) Plot of ${I_{\textrm{in},\epsilon}(|\psi\rangle)}/{I_{\text{Max}}}$ in terms of $n$ evolution steps  for $\sigma=10$, $p_+=1000$ and for different values of $m_0$. (b) Plot of ${I_{\textrm{in},\epsilon}(|\psi\rangle)}/{I_{\text{Max}}}$ in terms of $n$ evolution steps  for $m_0=1000$, $\sigma=10$ and for different values of $p_+$. (c) Plot of ${I_{\textrm{in},\epsilon}(|\psi\rangle)}/{I_{\text{Max}}}$ in terms of $n$ evolution steps  for $m_0=1000$, $p_+=1000$ and for different values of $\sigma$. The singularity correspond to the value $n\lambda=-100$ and is indicated by a black vertical line.}
	\label{fig:Figure3}
\end{figure}
We can then compute the mutual information between the ingoing and the $\epsilon$ degrees of freedom
\begin{equation}\label{I in e}
	I_{\textrm{in},\epsilon}(|\psi\rangle)=S(\rho^{\textrm{in}})+S(\rho^{\epsilon})-S(\rho^{\textrm{in},\epsilon})
\end{equation}
As in the previous subsection, it follows
\begin{align}
	I_{\textrm{in},\epsilon}(|\psi\rangle)&=\frac{1}{4}
	(1-4|D(\epsilon_1,\epsilon_2,0,0,n)|)\log\left(\frac{1}{4}-|D(\epsilon_1,\epsilon_2,0,0,n)|\right)
	\notag
	\\
	&+
	\frac{1}{4}(1+4|D(\epsilon_1,\epsilon_2,0,0,n)|)\log\left(\frac{1}{4}+|D(\epsilon_1,\epsilon_2,0,0,n)|\right)
	\notag
	\\
	&+
	\frac{1}{4}(1-4|D(\epsilon_1,\epsilon_2,j,j,n)|)\log\left(\frac{1}{4}-|D(\epsilon_1,\epsilon_2,j,j,n)|\right)
	\notag
	\\
	&+
	\frac{1}{4}(1+4|D(\epsilon_1,\epsilon_2,j,j,n)|)\log\left(\frac{1}{4}+|D(\epsilon_1,\epsilon_2,j,j,n)|\right)
	\notag
	\\
	&+
	\frac{1}{2}(2|D(\epsilon_1,\epsilon_2,0,0,n)+D(\epsilon_1,\epsilon_2,j,j,n)|-1)\log\left(\frac{1}{2}-|D(\epsilon_1,\epsilon_2,0,0,n)+D(\epsilon_1,\epsilon_2,j,j,n)|\right)
	\notag
	\\
	&-\frac{1}{2}(2|D(\epsilon_1,\epsilon_2,0,0,n)+D(\epsilon_1,\epsilon_2,j,j,n)|+1)\log\left(\frac{1}{2}+|D(\epsilon_1,\epsilon_2,0,0,n)+D(\epsilon_1,\epsilon_2,j,j,n)|\right)
	\notag
	\\
	&+\log(2) \ ,
\end{align}
which can be evaluated using \eqref{final D}. Results are reported in Figure \ref{fig:Figure3}. We see that, as the correlations between the outside and inside Hawking excitations decrease (Figure \ref{fig:Figure2}),  the correlations of the inside partner and the microscopic degrees of freedom increase as evolution brings it toward the strong quantum gravity regime close to the singularity. As in the previous subsection, the dependence on the various parameters is the expected one.

\subsubsection{Evolution of the correlations between the out going and the $\epsilon$ d.o.f}

In the previous subsection, we studied  the correlations between inside Hawking partner and the  microscopic $\epsilon$ sector (the one representing Planckian granularity in our toy model). Correlations grow as the partner falls into the singularity as its back reaction on the microscopic structure of the quantum geometry becomes more and more important as the singularity approaches.  A quantum counterpart of this is the fact that the outside partner escaping at infinity will correlate with the quantum degrees of freedom of the geometry as well even if it never comes into direct interaction with them (this is a consequence of the monogamy of entanglement). Indeed it is very easy to relate (in our system) the mutual information $I_{\textrm{in},\epsilon}(|\psi\rangle)$ with $I_{\textrm{out},\epsilon}(|\psi\rangle)$.  To see this one can first start from the definition of the mutual information between the subsystem corresponding to the outside partner and $\epsilon$, namely
\begin{equation}
I_{\textrm{out},\epsilon}(|\psi\rangle)=S(\rho^{\textrm{out}})+S(\rho^{\epsilon})-S(\rho^{\textrm{out},\epsilon}).
\end{equation} 
By comparing the density matrices \eqref{dm i} and \eqref{dm o} one obtains that
\begin{equation}
I_{\textrm{in},\epsilon}(|\psi\rangle)=I_{\textrm{out},\epsilon}(|\psi\rangle).
\end{equation}
Therefore, correlations between outside partner and the microscopic $\epsilon$ sector appear at the same rate as those between the inside partner and the $\epsilon$ sector.

\subsubsection{Interpretation}

The results in this section show consistently how the information coded in the initial correlations between the Hawking partners in $\sH_{in}\otimes \sH_{out}$ are transferred to the other degrees of freedom involved. This is expected to be the case due to the gravitational interaction between the matter sector and the spacetime geometry. What is particularly interesting, and novel in our case, is the generic transfer of entanglement to the `quantum hair' represented in our simple model by $\sH_{\epsilon}$. The computations show clearly, in the results reported in Figures \ref{fig:decoh}, \ref{fig:Figure2}, and \ref{fig:Figure3}, that correlations are established with these degrees of freedom during the evolution of the inside excitation toward the singularity. This indicates that, in a process where the mass of the black hole would be completely disappear via Hawking evaporation (a process that our simple model cannot account for), unitarity could be maintained via the purification of Hawking radiation granted by degrees of freedom like the $\epsilon$ sector of this toy model.   Despite the simplistic nature of the model, we notice that it complies with the expected behaviour of the back reaction: the lower the mass (the smaller $m_0$), or equivalently the smaller the black hole, the stronger the development of entanglement between the matter and the defects in the quantum geometry (stronger interaction with the Planckian granularity). The effect also grows with the particle momentum ${\rm p}_{\phi}$ and dispersion $\sigma$, as expected.

\section{Discussion}\label{disc}

The problem of unitarity in black hole formation and evaporation necessitates a deep understanding of the strong quantum gravity regime near the singularity beyond the black hole horizon. This presents a challenge for both the {\em holographic} and {\em non-holographic} perspectives. In the holographic view, macroscopic black holes are believed to possess a finite number of internal degrees of freedom bounded by their horizon area. Consequently, they are expected to follow a standard Page curve, wherein the exterior entropy starts decreasing at the macroscopic stage known as the Page time. Conversely, in the non-holographic perspective, which is the framework of this work, the number of internal degrees of freedom inside black holes is unrelated to their horizon area. In this scenario, the exterior entropy continues to grow while the black hole is macroscopic, and the purification occurs after the semiclassical regime.
In non-holographic scenarios, where information is recovered after the complete evaporation of a black hole, it becomes essential to identify the appropriate purifying degrees of freedom. Subsequently, understanding how the correlations between the Hawking partners are dynamically transferred to these degrees of freedom becomes crucial.

Suitable degrees of freedom are those that can persist after complete evaporation without significantly affecting the energy budget, considering that only a Planck mass energy is available for the multitude of degrees of freedom needed to purify the emitted Hawking radiation until the black hole's total evaporation.
Approaches to quantum gravity, which postulate a discrete fundamental nature of matter and geometry at the Planck scale, offer a natural set of such degrees of freedom. In these theories, the concepts of smooth geometry and continuous fields are approximate and emerge only when probed with sufficiently coarse measurements. Within this framework, flat configurations such as Minkowski spacetime with the Minkowski vacuum for matter fields are expected to exhibit high degeneracy in a coarse-grained sense, meaning that a single macroscopic configuration corresponds to a large number of microscopic states. Any deviations among these microscopic states can be viewed as defects in the fundamental structure that are indistinguishable at the macroscopic level. These precisely are the types of degrees of freedom suitable for purification. They are numerous (due to their local nature and increasing number with spatial extension) and, in the aforementioned sense, possess negligible weight or influence on the system's overall energy.

Once suitable degrees of freedom are identified, one needs to understand the dynamical process by which unitarity is preserved when these are tracked during the black hole formation and evaporation.  In particular, while the black hole is macroscopic and emits Hawking radiation, one needs to understand the mechanism by which the (initially) maximally correlated Hawking pairs evolve into a particle escaping to infinity as Hawking radiation correlated to the Planckian degrees of freedom  inside that will emerge (after complete evaporation) as the purifying defects of the previous paragraph.

Both ingredients are shown to be present in the simple model of quantum black hole analysed in this work. 
The model is expected to capture the dynamics of an infalling scalar excitation when it is spherically symmetric
in the near singularity approximation. It can be used to model the near singularity dynamics of a Hawking pair of scalar particles including the back reaction of the
inside partner as it approaches the near singularity regime.   
On the one hand the  toy model, used here, mimics the quantization of gravity in the framework of loop quantum gravity and thus 
contain the type of microscopic `weightless' degrees of freedom. More precisely,  the spectrum of the mass operator is infinitely degenerate, due to the existence of quantum hair associated to (what we call) the $\epsilon$ sectors, Section \ref{degeM}.
On the other hand, we showed explicitly how correlations are transferred dynamically to the `hidden' quantum hair as the
inside excitation approaches the near singularity regime. This was done first by using decoherence function (suitable intuitive method for the simplified setup of the toy model because there is a single such function), and later using the basis independent mutual information entanglement measure.  

Of course, the model does not explain how the black hole evaporates and how the information remains after complete evaporation coded in the fundamental `hair'. This would possibly require making the extremely non trivial step of allowing local degrees of freedom and treating their strong quantum dynamics near the singularity. We are not sure if one can find intermediate models between our simplistic one and the fully general situation that requires treatment without any possible approximation. Yet, even when one is still far from the full understanding of the process of black hole formation and evaporation (in approaches like loop quantum gravity and others), we hope that the indications provided by models, like the one studied here, stimulate possibilities and new ideas that could show useful for researchers investigating this central problem.

\section{Acknowledgement}

We are grateful for the hospitality of the {\em John Bell Institute}, and we thank the participants of the workshop the {\em Black Hole Information Puzzle} in 2022 for inputs and discussions. This publication was made possible through the support of the ID\# 62312 grant from the John Templeton Foundation, as part of the \href{https://www.templeton.org/grant/the-quantuminformation-structure-ofspacetime-qiss-second-phase}{`The Quantum Information Structure of Spacetime' Project (QISS)}. The opinions expressed in this project/publication are those of the author(s) and do not necessarily reflect the views of the John Templeton Foundation.

\providecommand{\href}[2]{#2}\begingroup\raggedright\endgroup

\end{document}